\documentclass[3p,number,sort&compress]{elsarticle}

\pdfoutput=1
\usepackage{dcolumn}
\usepackage{bm}
\usepackage{amsmath}
\usepackage{amsthm}

\usepackage{amsfonts}
\usepackage{euscript,bbm}
\usepackage{ifthen}
\usepackage{psfrag}
\usepackage{slashed}
\usepackage{hyperref}
\usepackage{textcomp,gensymb}
\usepackage{calc}
\usepackage{color}
\usepackage{lineno}

\usepackage{marvosym}

\newcommand{\cns}{CE\ensuremath{\nu}NS}

\newcommand{\be}{\begin{equation}}
\newcommand{\ee}{\end{equation}}
\newcommand{\bea}{\begin{eqnarray}}
\newcommand{\eea}{\end{eqnarray}}

\journal{Nuclear Instruments and Methods A}

\begin{document}
\begin{frontmatter}

\title{Background Studies for the MINER Coherent Neutrino Scattering Reactor Experiment}

\author[a]{G.~Agnolet} 
\author[a]{W.~Baker} 
\author[b]{D.~Barker}
\author[a]{R.~Beck}
\author[c]{T.J.~Carroll} 
\author[c]{J.~Cesar}
\author[b]{P.~Cushman}
\author[d]{J.B.~Dent}
\author[c]{S.~De~Rijck} 
\author[a]{B.~Dutta} 
\author[c]{W.~Flanagan} 
\author[b]{M.~Fritts}
\author[a,e]{Y.~Gao}
\author[a]{H.R.~Harris}
\author[a]{C.C.~Hays} 
\author[f]{V.~Iyer}
\author[a]{A.~Jastram}
\author[a]{F.~Kadribasic}
\author[b]{A.~Kennedy} 
\author[a]{A.~Kubik}
\author[g]{I.~Ogawa} 
\author[c]{K.~Lang} 
\author[a]{R.~Mahapatra}
\author[b]{V.~Mandic}
\author[h]{R.D.~Martin}
\author[b]{N.~Mast}
\author[i]{S.~McDeavitt}
\author[a]{N.~Mirabolfathi} 
\author[f]{B.~Mohanty}
\author[g]{K.~Nakajima} 
\author[i]{J.~Newhouse}
\author[j]{J.L.~Newstead}
\author[c]{D.~Phan}
\author[c]{M.~Proga}
\author[k]{A.~Roberts}
\author[l]{G.~Rogachev}
\author[c]{R.~Salazar} 
\author[k]{J.~Sander} 
\author[f]{K.~Senapati}
\author[g]{M.~Shimada} 
\author[a]{L.~Strigari} 
\author[g]{Y.~Tamagawa} 
\author[a]{W.~Teizer} 
\author[i]{J.I.C.~Vermaak}
\author[b]{A.N.~Villano} 
\author[m]{J.~Walker}
\author[a]{B.~Webb}  
\author[a]{Z.~Wetzel}
\author[c]{S.A.~Yadavalli}

\address[a]{Department of Physics and Astronomy, and the Mitchell Institute for Fundamental Physics and Astronomy, 
Texas A\&M University, College Station, TX 77843, USA }
\address[b]{School of Physics \& Astronomy, University of Minnesota, Minneapolis, MN 55455, USA} 
\address[c]{Department of Physics, University of Texas at Austin, Austin, TX 78712, USA}
\address[d]{Department of Physics, University of Louisiana at Lafayette, Lafayette, LA 70504, USA}
\address[e]{Department of Physics \& Astronomy, Wayne State University, Detroit, MI 48201, USA}
\address[f]{School of Physical Sciences, National Institute of Science Education and Research, Jatni - 752050, India}
\address[g]{Graduate School of Engineering, University of Fukui, Fukui, 910-8507, Japan}
\address[h]{Department of Physics, Engineering Physics \& Astronomy, Queen's University, Kingston, Ontario, Canada}
\address[i]{TEES Nuclear Science Center, Texas A\&M University, College Station, TX 77843, USA}
\address[j]{Department of Physics, Arizona State University, Tempe, AZ 85287, USA}
\address[k]{Department of Physics, University of South Dakota, Vermillion, SD 57069, USA}
\address[l]{Cyclotron Institute, Texas A\&M University, College Station, TX 77843, USA}
\address[m]{Department of Physics, Sam Houston State University, Huntsville, TX 77341, USA}

\begin{abstract}
The proposed Mitchell Institute Neutrino Experiment at Reactor (MINER) experiment at the Nuclear Science Center at Texas A\&M University will search for 
coherent elastic neutrino-nucleus scattering within close proximity (about 2 meters) of a 1\,MW TRIGA nuclear reactor core using 
low threshold, cryogenic germanium and silicon detectors.   
Given the Standard Model cross section of the scattering process and the proposed experimental proximity to the reactor, as many as 
5 to 20\,events/kg/day are expected.  We discuss the status of preliminary measurements to characterize the main backgrounds for 
the proposed experiment.  Both \textit{in situ} measurements at the experimental site and 
simulations using the MCNP and GEANT4 codes are described.  A strategy for monitoring backgrounds during data taking is briefly discussed.

\end{abstract}
\end{frontmatter}

\section{Introduction} 
\label{sec:intro}

The cross section for the coherent elastic scattering of neutrinos off of nuclei (\cns)~\cite{Freedman:1973yd} is a long-standing prediction of the Standard Model, but has yet to be measured 
experimentally in part due to the extremely low energy threshold needed for detection with typical high flux neutrino sources such as nuclear reactors.  Improvements 
in semiconductor detector technologies~\cite{Mirabolfathi:2015pha} 
which utilize the Neganov-Luke phonon amplification method~\cite{Luke:1990ir}     
have brought {\cns} detection within reach.  The Mitchell Institute Neutrino Experiment at Reactor (MINER) experiment, currently under development at the 
Nuclear Science Center (NSC) at Texas A\&M University, 
will leverage this detector technology to detect {\cns} and measure its cross section.  If successful, the {\cns} interactions can be used to 
probe new physics scenarios including a search for sterile neutrino oscillations, the neutrino magnetic moment, and other processes beyond the 
Standard Model~\cite{Barranco:2005yy,Scholberg:2005qs,Dutta:2015nlo,Dutta:2015vwa}.  
The experiment will utilize a megawatt-class 
TRIGA (Training, Research, Isotopes, General Atomics) pool reactor stocked with low-enriched (about\,20\%) $^{235}$U.  
This reactor has an unique advantage of having a movable core and provides access to deploy detectors as close as about 1\,m from the reactor, allowing for a varying 
distance from the neutrino source to the detector.  At these short baselines, and given 
the Standard Model cross section, we expect to detect as many as 5 to 20\,events/kg/day in the range of recoil energy between 
10 and 1000~eV$_{\rm nr}$.

An important aspect of the proposed experiment are the backgrounds induced by both the core and 
environmental sources.  These backgrounds include gammas and neutrons from the reactor, muons and muon-induced 
neutrons from cosmic rays, and ambient gammas.  The rate of such backgrounds must be comparable to or below 
the expected rate of the neutrino recoil signal.  We take a rate of 100\,events/kg/day in the range of recoil energy between 
10 and 1000~eV$_{\rm nr}$ as the target level of acceptable background rate, corresponding to a signal to background ratio of about 0.1 to 0.2.  Events outside of this energy window 
are acceptable to a reasonable level (about 100\,Hz total event rate) and can serve to normalize backgrounds in the signal region.

The paper is organized as follows.  In Section~\ref{sec:expHall}, a brief description of the experimental location is given.  Section~\ref{sec:bkgSim} describes the modeling of the 
reactor core and experiment in the MCNP and GEANT4 framework.  Sections~\ref{sec:Gamma}, \ref{sec:Neutron}, and \ref{sec:Muon} describe the \textit{in situ} 
measurements of the gamma, neutron, and cosmic muon backgrounds respectively, including comparison to the simulation for the gamma and neutron 
backgrounds.  Section~\ref{sec:rateEstimate} combines the simulation with the \textit{in situ} measurements 
to estimate a background rate in the detectors given a preliminary shielding design.  Finally, status and prospects are described in Section~\ref{sec:summary}.

\section{Description of Experimental Site}
\label{sec:expHall}

\begin{figure}[ht]
\centering
\hspace{-5pt}
\includegraphics[width=6.6in]{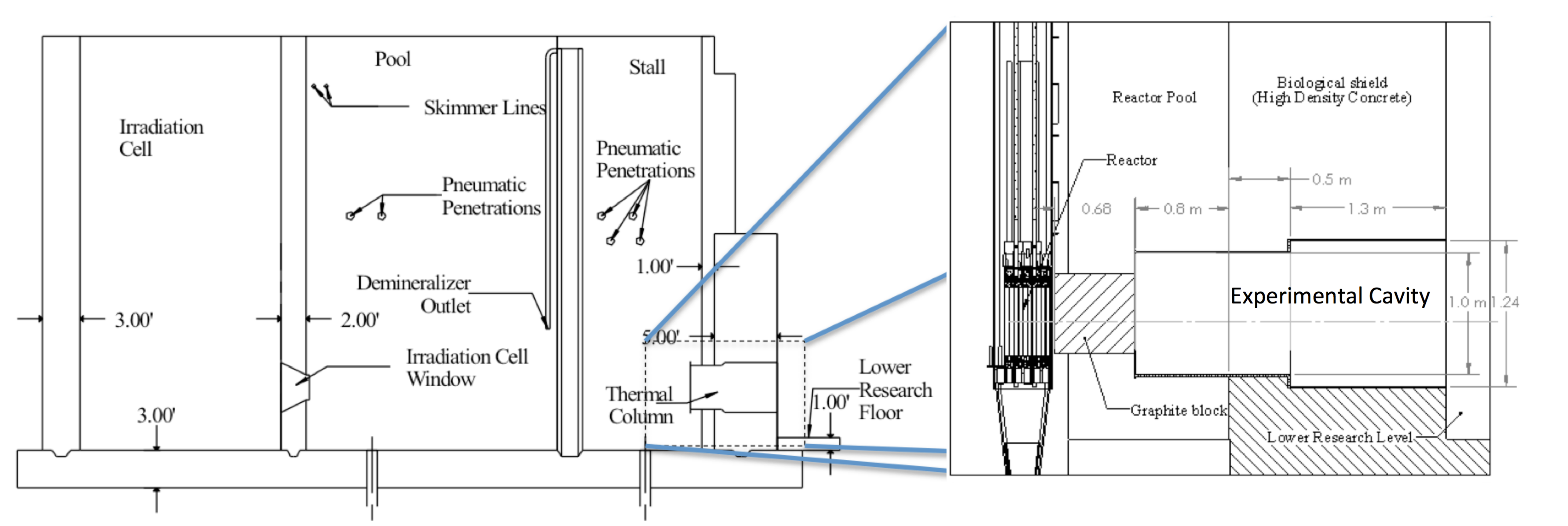}
\caption{
\footnotesize Schematic side view of the reactor pool and experimental cavity where the proposed detector and shielding will be constructed. }
\label{fig:TCdiagram}
\end{figure}

\begin{figure}[ht]
\centering
\hspace{-5pt}
\includegraphics[height=2.1in]{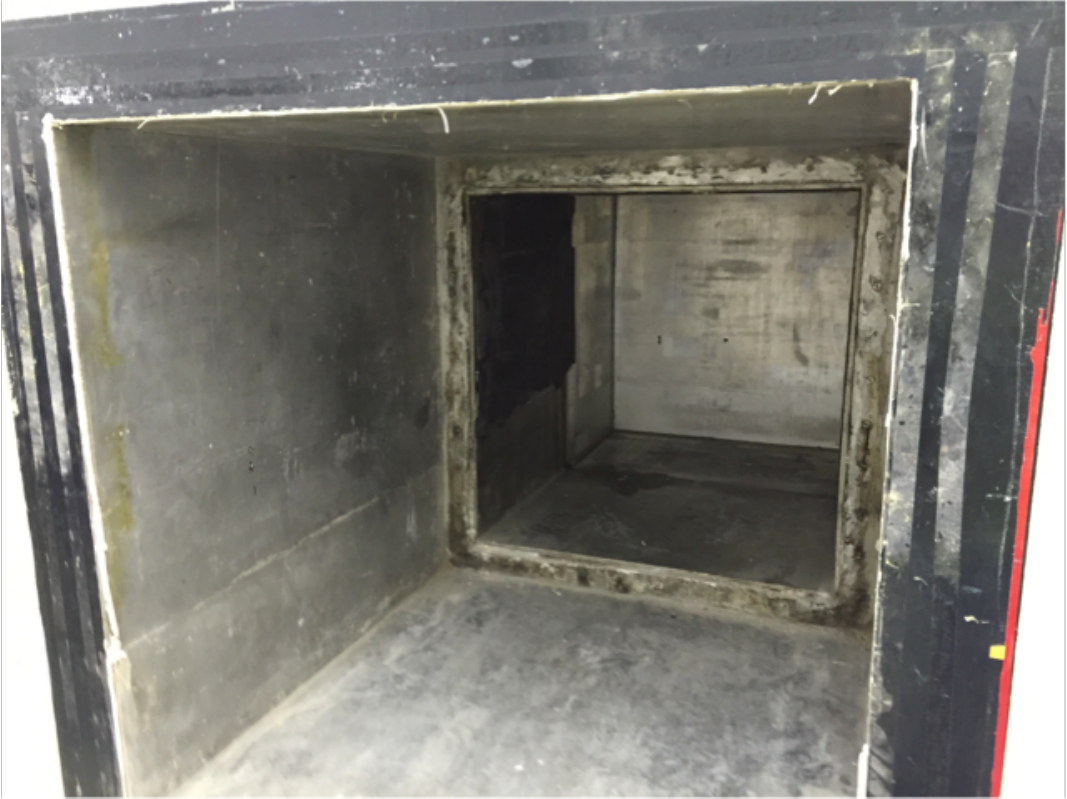}
\caption{
\footnotesize Photograph from the outside of the empty experimental cavity. }
\label{fig:TCphoto}
\end{figure}

The NSC reactor facility pool is surrounded by roughly 2 meters of high density concrete (about 3.5\,g/cm$^3$ density) which acts as a shield to the 
high flux of neutron and gamma byproducts in the reactor.   A cavity in this wall, dubbed the ``Thermal Column", was used in the past to 
facilitate close proximity to the reactor for material neutron irradiation.  The cavity is located in the lower research area of the NSC and is in the same horizontal 
plane as the reactor core (see Figure~\ref{fig:TCdiagram}).  The cavity has many advantages as an experimental location, including 
the ability to access an area in very close proximity to the core, a natural overburden provided by the concrete wall to reduce the rate of cosmic muons, and an
open area to allow placement of optimized shielding between the core and the detectors.  A schematic diagram of the Thermal Column 
can be seen in Figure~\ref{fig:TCdiagram} and a photograph from the outside of the cavity is shown in Figure~\ref{fig:TCphoto}.

\section{Background Simulation}
\label{sec:bkgSim}

\subsection{Reactor Core Model}
\label{sec:coreModel}

\begin{figure}[ht]
\centering
\hspace{-5pt}
\includegraphics[height=2.1in]{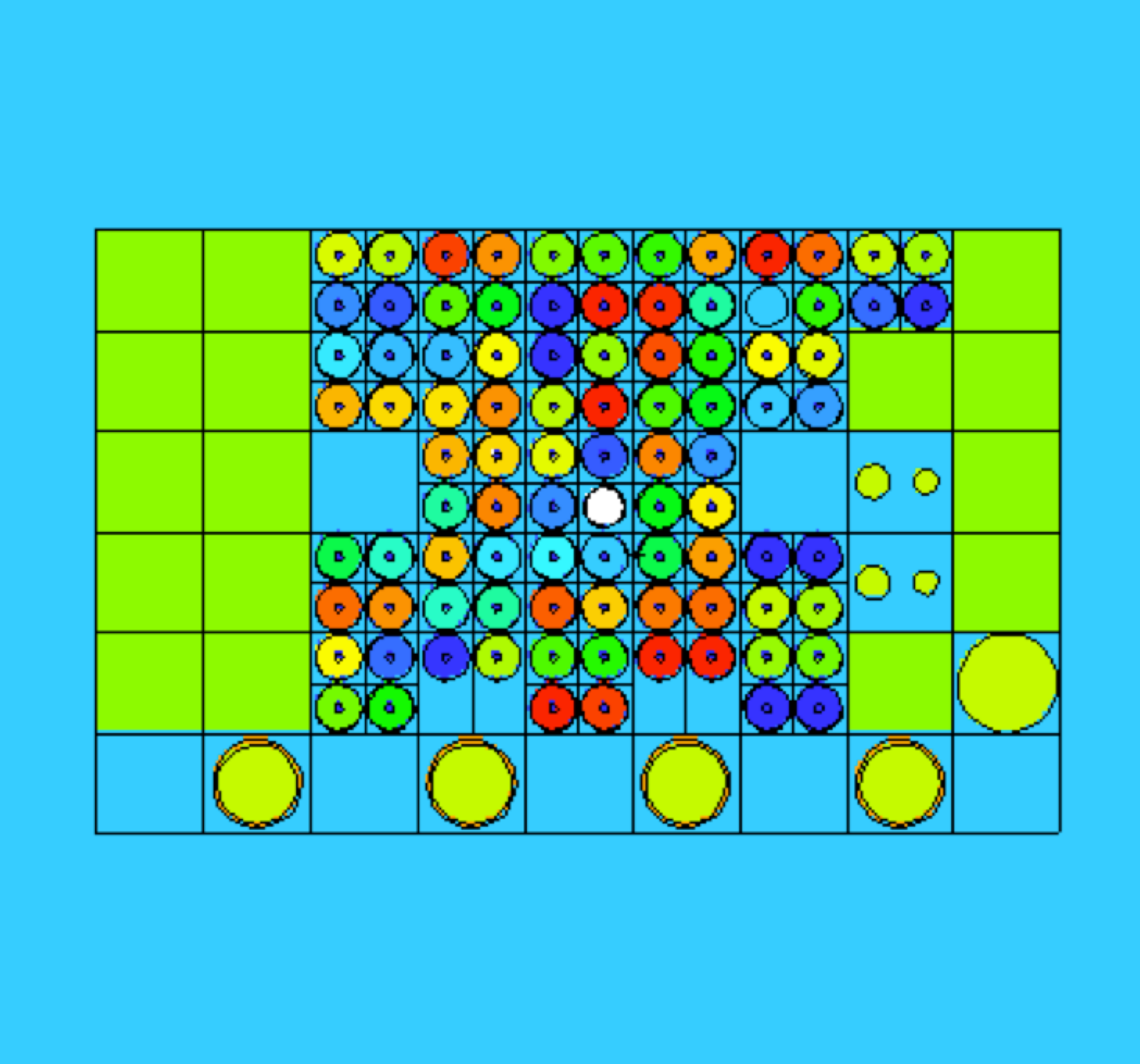}
\includegraphics[height=2.1in]{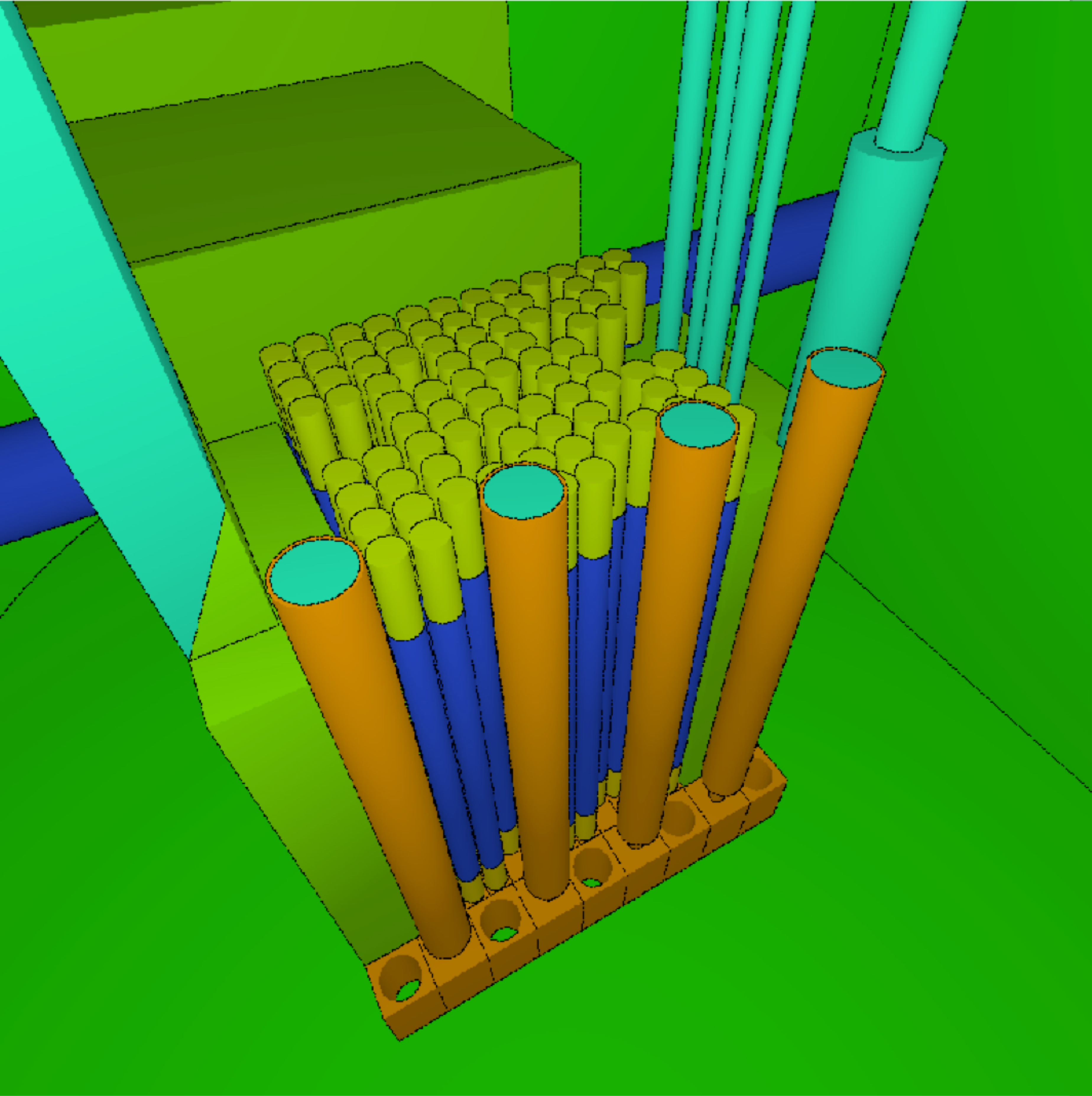}
\caption{
\footnotesize Visualization of the TRIGA reactor core as modeled in MCNP.  The small circles in the picture represent the various fuel and control rods while the green 
rectangles are graphite reflectors.  The fuel rods are about 3.58\,cm in diameter and 38.1\,cm in length.}
\label{figure:MCNPCoreModel}
\end{figure}

Fission processes in the reactor produce large fluxes of both gammas and neutrons near the core.  
The energy spectrum and production rate of these backgrounds are predicted using a core model developed at the NSC, shown 
in Figure~\ref{figure:MCNPCoreModel}, and applied in the MCNP~\cite{mcnp} framework. The TRIGA reactor of the NSC features a 90 fuel element, 
low-enriched uranium core operating at a nominal power of 1\,MW. The fuel burn-up of the relatively new core (installed in 2006) is modeled in a 15 axial layer 
configuration for each fuel element and includes a wide range of fission products in the fuel material resulting in a high detail model of the reactor.

\begin{figure}[ht]
\centering
\hspace{-5pt}
\includegraphics[width=3.0in]{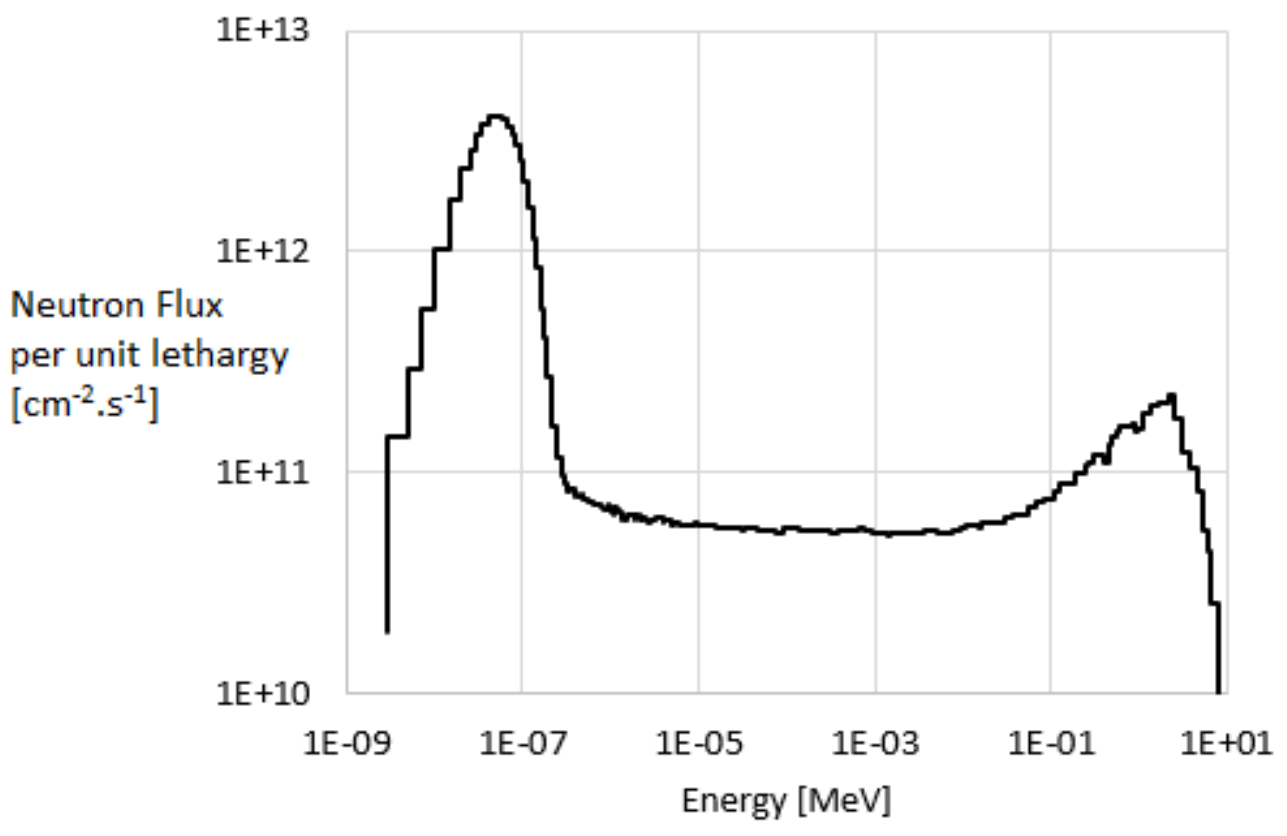}
\includegraphics[width=3.0in]{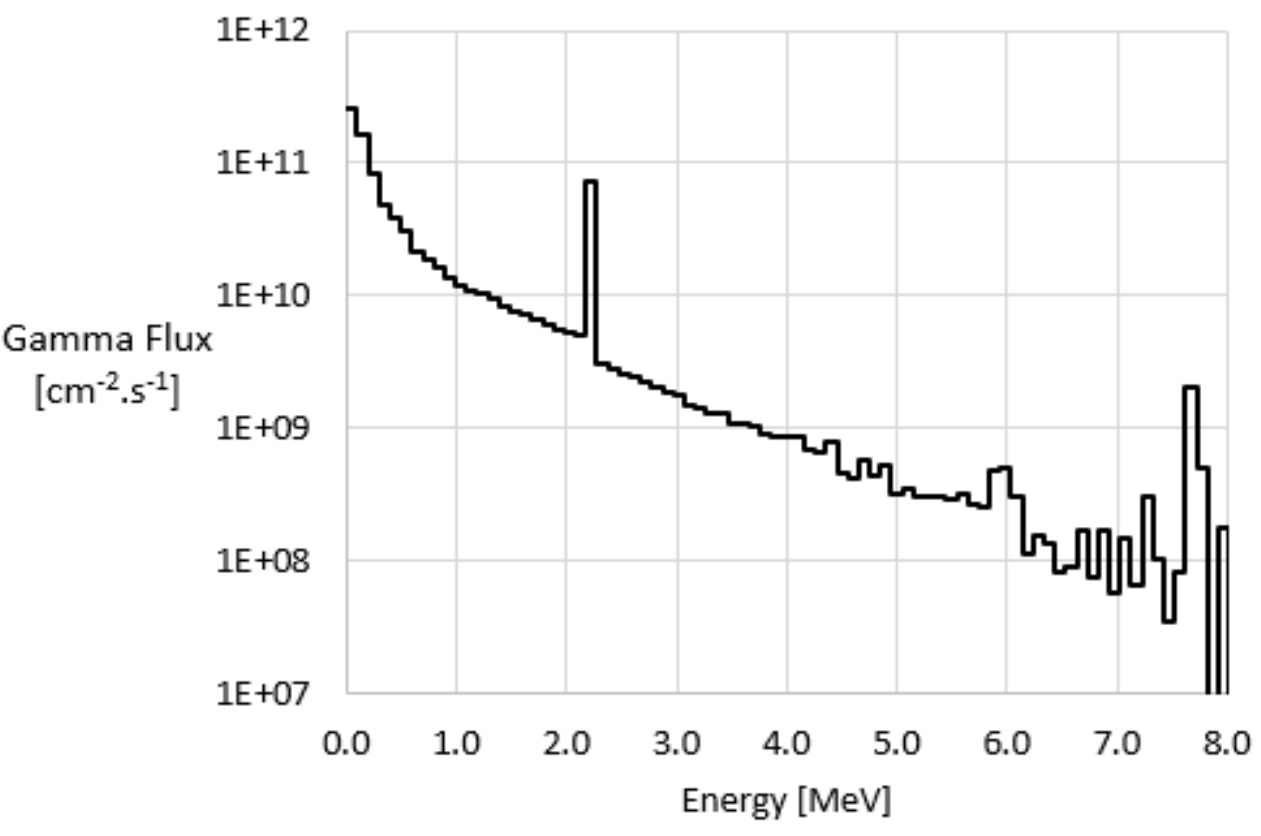}
\caption{
\footnotesize Calculated neutron (left) and gamma (right) spectra just outside the core volume (before 
shielding), obtained using the MCNP core model as described in the text.  The neutron spectrum is bin-by-bin normalized to $E_{Ave}/(E_2 - E_1)$ where $E_1$ and $E_2$ are 
the energy values at the respective bin edges and $E_{Ave}$ is the average of these bin edge values.  The 2.2 MeV line in the gamma spectrum is the result 
of neutron capture on Hydrogen in the surrounding water, while excesses in bins near 6 and 8 MeV are due to statistical fluctuations.  The 
total fluxes from these calculations are $5.8\times\,10^{11}$ cm$^{-2}$~s$^{-1}$ fast neutrons 
component ($> 100$ keV kinetic energy), $7.7\times 10^{12}$\,cm$^{-2}$~s$^{-1}$
 thermal neutrons ($< 0.625$ eV kinetic energy), and $9.0\times 10^{11}$\,cm$^{-2}$~s$^{-1}$ gammas.}
\label{figure:MCNPSpectrum}
\end{figure}

Using the MCNP reactor core model, we calculated the neutron energy spectrum produced by the reactor shown on the left in Figure~\ref{figure:MCNPSpectrum}, 
with fluxes of $5.8\times\,10^{11}$ cm$^{-2}$~s$^{-1}$ fast 
component ($> 100$ keV kinetic energy) and $7.7\times 10^{12}$\,cm$^{-2}$~s$^{-1}$
 thermal component ($< 0.625$ eV kinetic energy).  A moderator can be used to convert the fast neutron flux to a thermal flux which can 
 be shielded using a thermal neutron absorber such as boron, cadmium or gadolinium.  
 
The simulated gamma spectrum is shown on the right in Figure~\ref{figure:MCNPSpectrum}, with a total flux of $9.0\times 10^{11}$\,cm$^{-2}$~s$^{-1}$. This gamma energy spectrum can be 
attenuated by conventional high density materials such as lead.

\subsection{GEANT4 Geometry Model}
\label{sec:g4setup}

A model geometry of the experimental hall was constructed in 
the GEANT4~\cite{Agostinelli:2002hh} (v10.2.1) framework.  Correct description of the atomic composition of the surrounding materials 
is critical since backgrounds strongly depend on the materials used and secondary production of backgrounds in these materials must be included.  Detailed 
material descriptions with isotope composition to the level of ppm are provided by the NSC and are included in the GEANT4 model.  

Reactor gammas and neutrons were generated with an energy spectrum as 
produced by the MCNP core model described above.  The flux is modeled to originate from a $30\times30$\,cm$^2$ square plane representing the active face of the reactor 
closest to the experimental cavity.  To model different core positions, this 
source surface was moved to the corresponding core face position.  To significantly save computational time, particles were simulated with only momenta along a direction
from the reactor core to the experimental cavity, perpendicular to the source plane.  The ``Shielding"~\cite{shielding} physics list was used in all GEANT4 simulations 
described in this paper.

Since the initial flux from the reactor is large and the desired target rate at the detectors must be low, we implemented a variance reduction scheme 
to obtain good statistical significance of the background characteristics at the detector site.  GEANT4 has multiple built-in variance reduction schemes 
available to users, and we chose the `importance sampling' scheme for this application.  In the importance sampling scheme, the geometry is divided up into different regions, each with an 
importance score assigned.  As a particle is propagated across the boundary of two such regions, the ratio of the importance score in the new region over that of the previous region is taken.  If this ratio 
is larger than one, the particle is duplicated a number of times equal to this ratio decreased by one.  If less than one, a ``Russian Roulette" algorithm is used to determine whether to terminate the 
particle with a probability equal to the ratio.  Particles are then weighted by the inverse of the importance ratio.  By increasing the importance value assigned to 
regions deeper within the shielding, the number of particles making it through to the detector  
is greatly enhanced. We used 16 to 22 importance regions (depending on distance of the source to the detectors) of equal thickness with importance score increasing by factors of 2.  
Large statistics GEANT4 simulations (typically around $10^9$ primary particles generated) were run on both the Brazos Computing Cluster at Texas A\&M University as well as the 
Texas Advanced Computing Center (TACC) cluster at the University of Texas at Austin.

\begin{figure}[ht]
\centering
\hspace{-5pt}
\includegraphics[width=3.5in]{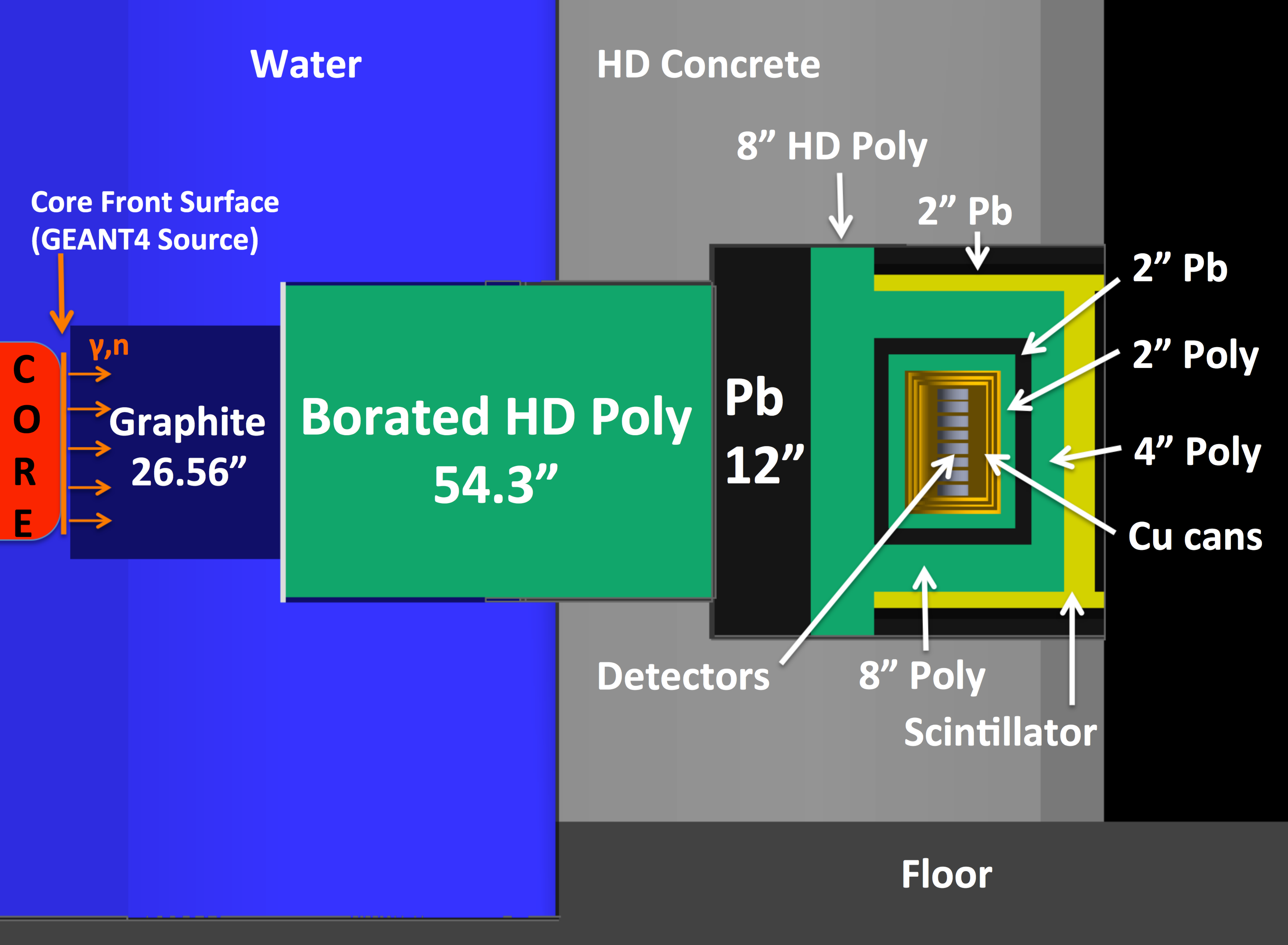}
\caption{
\footnotesize Cross section of the GEANT4 geometry model of the experimental cavity with a preliminary shield design.  In this figure, different materials are represented by 
different colors.  Blue: water, Dark Blue: graphite, Green: 5\% Borated polyethylene, Light Gray: HD concrete, Dark Gray: Lead, Gold: Copper, Yellow: Plastic Scintillator.  The 
small disks within the copper are germanium and silicon detectors.}
\label{fig:ShieldDesign1}
\end{figure}

A preliminary shielding design was then added to the GEANT4 geometry model to assess the background expected in the full experimental 
setup.  Materials included 1.38~m of high density borated (5\%) polyethylene as neutron shielding and 30.5~cm of lead as gamma shielding.  Additional 
lead and polyethylene were included after the initial shielding to reduce backgrounds from secondary particles which include 
neutrons from ($\gamma$,n) reactions and gammas released from neutron captures in the shielding materials.  
This shielding design is shown in Figure~\ref{fig:ShieldDesign1}.  The thickness of shielding materials in this design was chosen based on initial estimates made with 
simple GEANT4 geometry models but has not yet been optimized.

\section{Gamma Background Measurements}
\label{sec:Gamma}

\begin{figure}[ht]
\centering
\hspace{-5pt}
\includegraphics[height=2.1in]{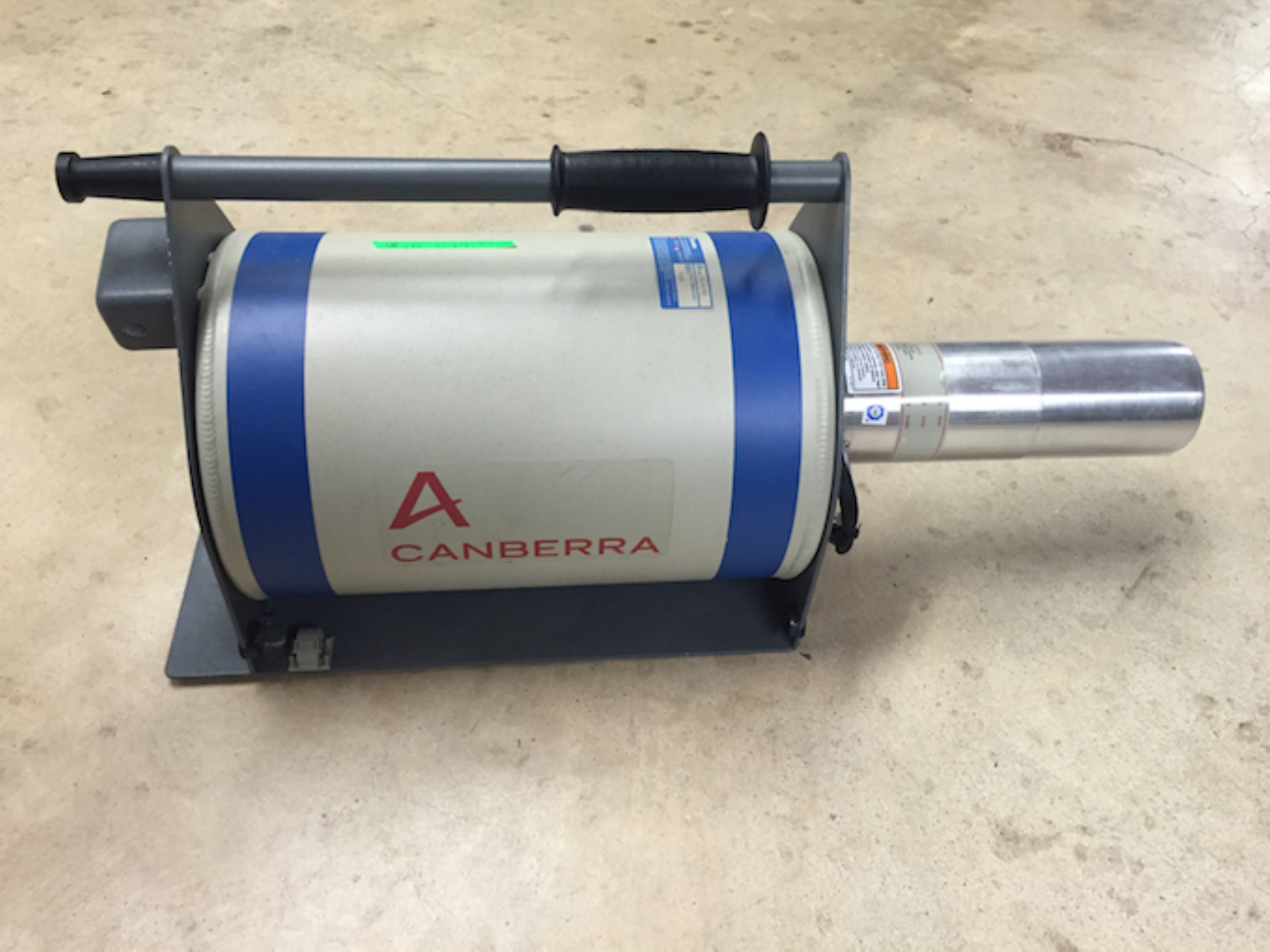}
\includegraphics[height=2.1in]{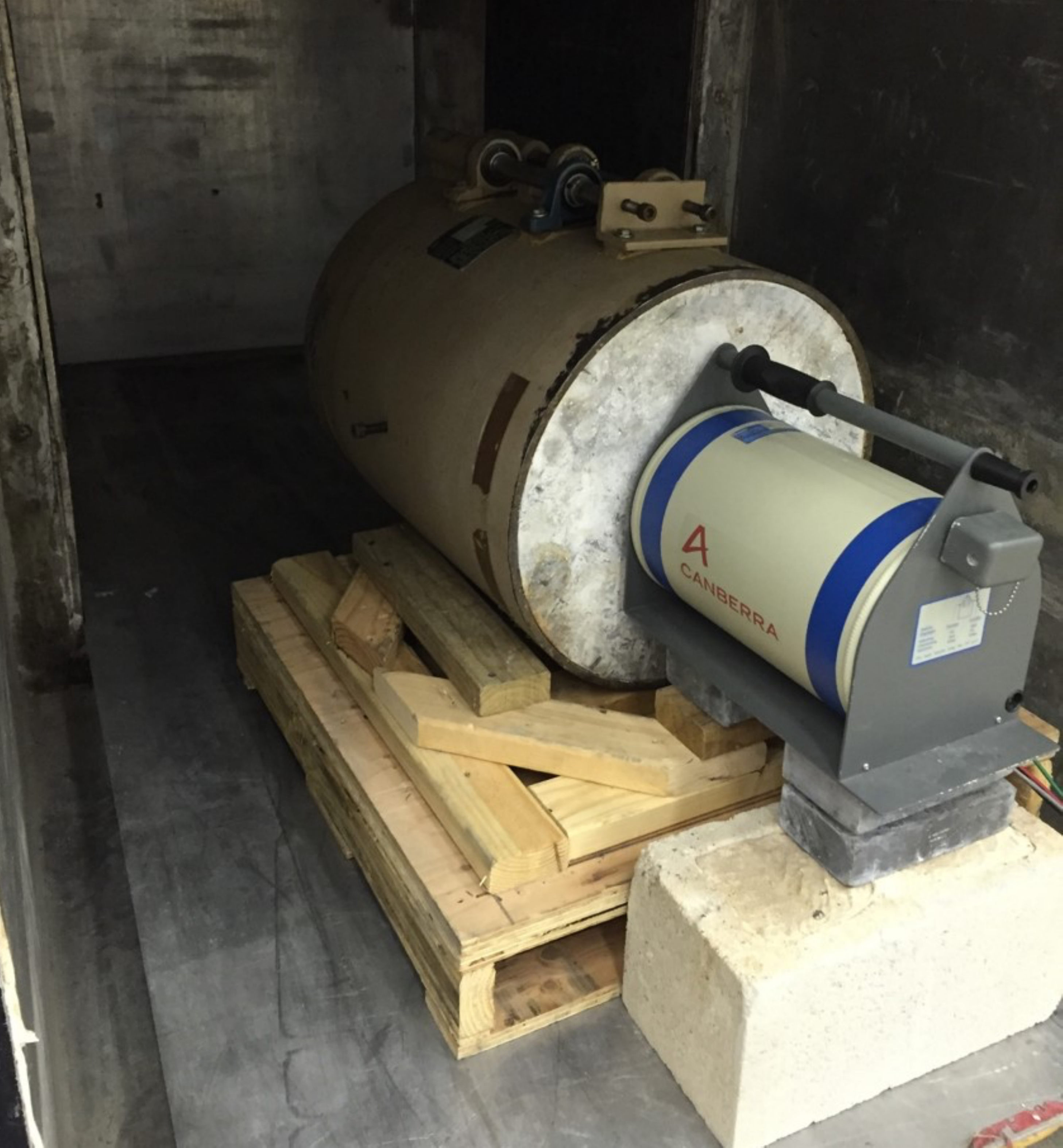}
\includegraphics[width=3.0in]{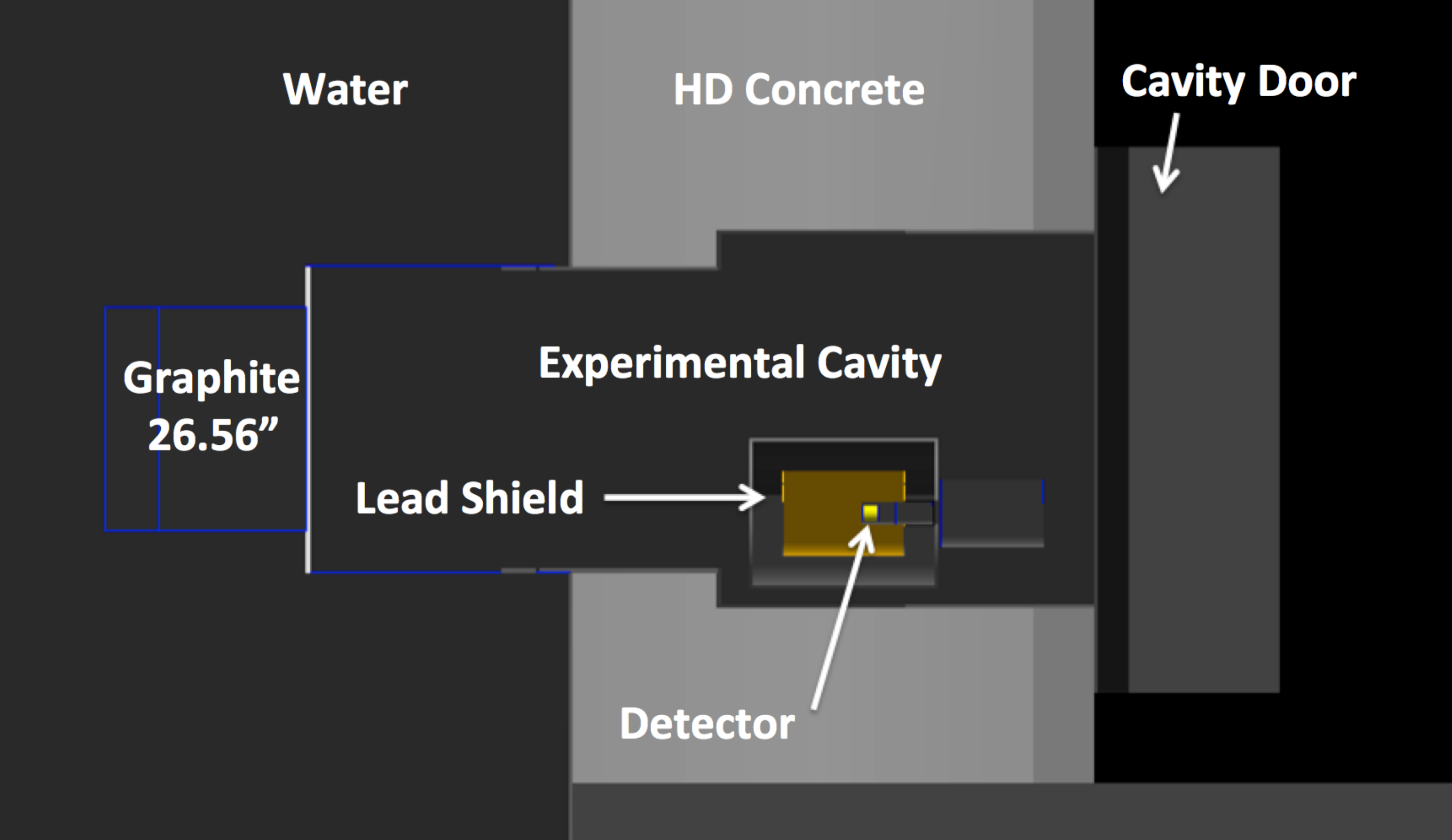}
\caption{
\footnotesize  Top Left: Photo of Canberra HPGe detector used in the gamma measurement. Top Right: Placement of HPGe detector and shielding in the experimental 
cavity for the gamma measurement.  The 
active detector is shielded by approximately 4 inches of lead provided by the commercial shield described in the text.  Bottom: Cross section of GEANT4 geometry (side view) used in simulation of 
this setup.  The active detector is shown in yellow.  The only shielding present in this configuration is the commercial gamma shielding described in the text (the experimental cavity was 
otherwise empty). }
\label{fig:Geant4HPGe}
\end{figure}

\begin{figure}[ht]
\centering
\hspace{-5pt}
\includegraphics[width=3.0in]{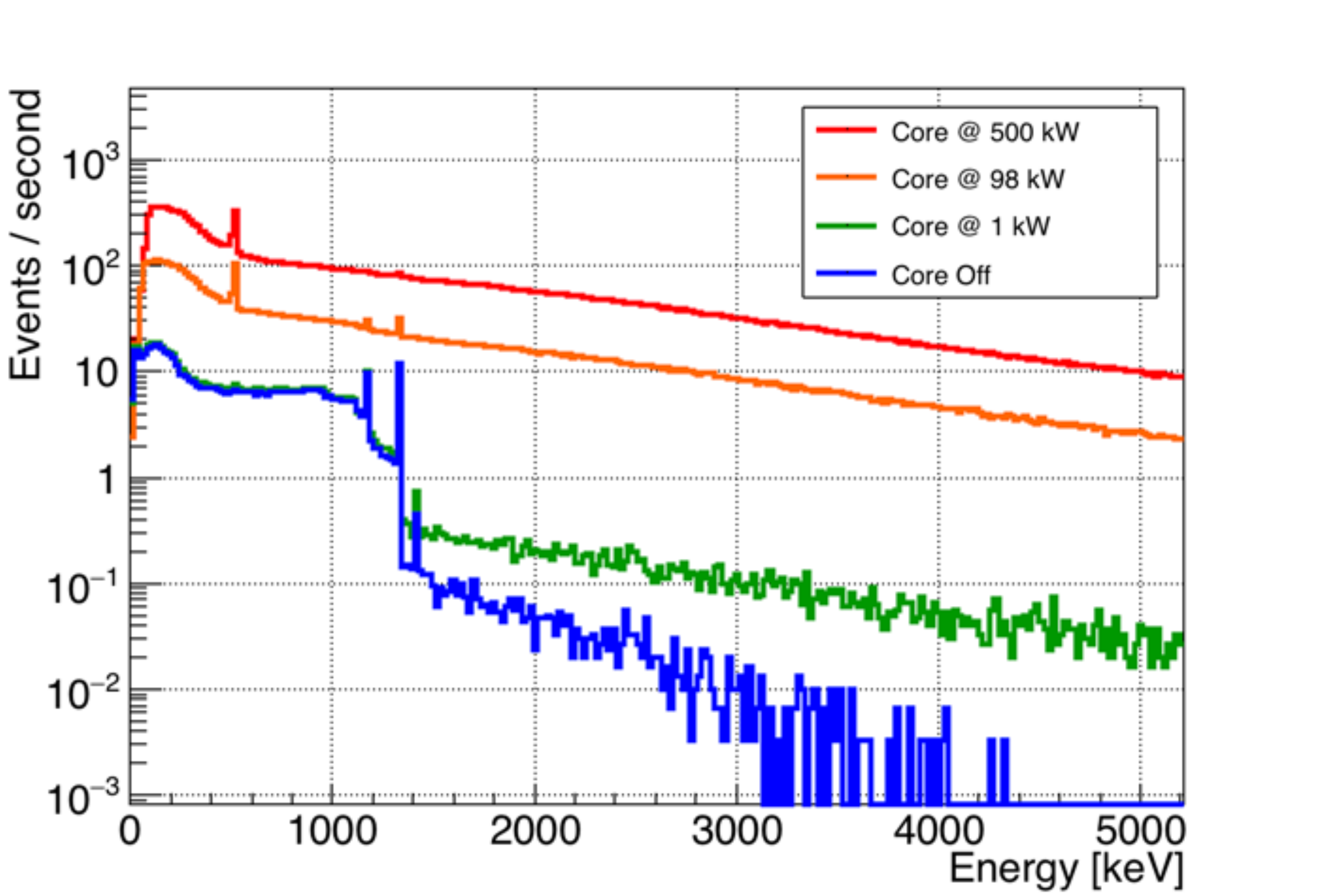}
\includegraphics[width=3.0in]{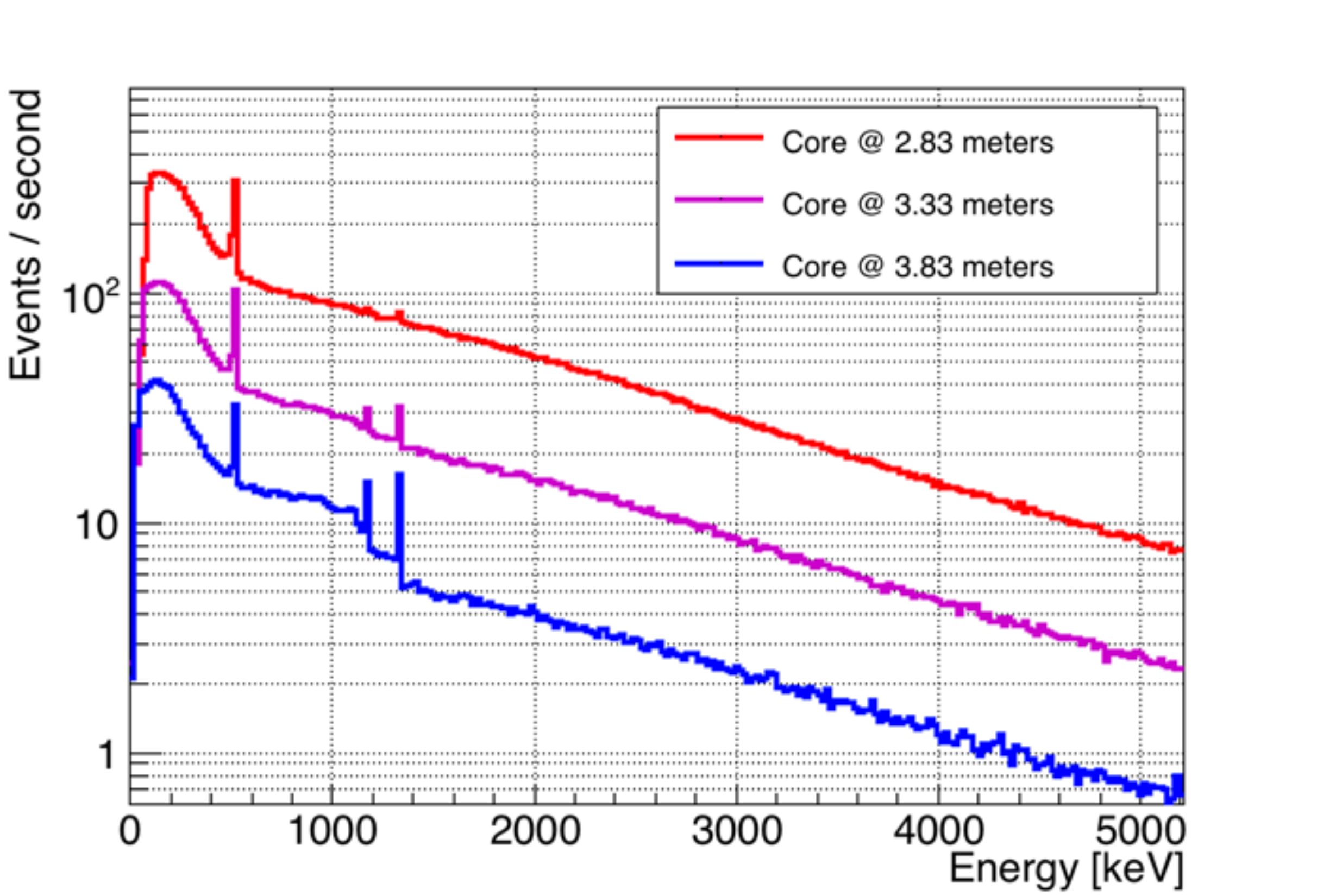}
\caption{
\footnotesize Measured gamma spectrum in the commercial HPGe detector (Canberra GC2020, approx. 0.5kg) for different reactor powers with the core held at 3.33 meters (left) 
and different reactor positions with the core at 98 kW power (right).  All runs consist of 300 seconds of live time. }
\label{figure:MeasuredGammaSpectrum}
\end{figure}

Background measurements have been conducted in the experimental cavity using a commercial High Purity Germanium (HPGe) detector shown in Figure~\ref{fig:Geant4HPGe} 
(Canberra GC2020, approx. 0.5\,kg).   Due to the large volume of 
water between the detector and the reactor core, these measurements were dominated by gamma interactions. A commercially available shield was used to limit the 
rate registered by the detector, while maintaining a simple geometry for matching with simulations. The shield is cylindrical, comprised of a 4" layer of low activity lead enclosed externally by 
a 1/2" thick layer of steel. Inside the shield cavity, the lead is lined with layers of high purity tin and OFHC copper, approximately 0.04" and 0.06" thick, respectively, to block lead X-rays. Measurements 
were made at reactor powers of 0\,kW (core off), 1\,kW, 98\,kW, and 500\,kW at distances (measured from the face of the core to the face of the experimental cavity) of 3.83\,m, 3.33\,m, and 2.83\,m 
(see Figure~\ref{figure:MeasuredGammaSpectrum}).  All measurements consist of 300 seconds of live time.  The HPGe detector 
was calibrated before measurements using a $^{22}$Na source, which provides gammas with energy 511\,keV and 1274\,keV.

\begin{figure}[ht]
\centering
\hspace{-5pt}
\includegraphics[width=3.0in]{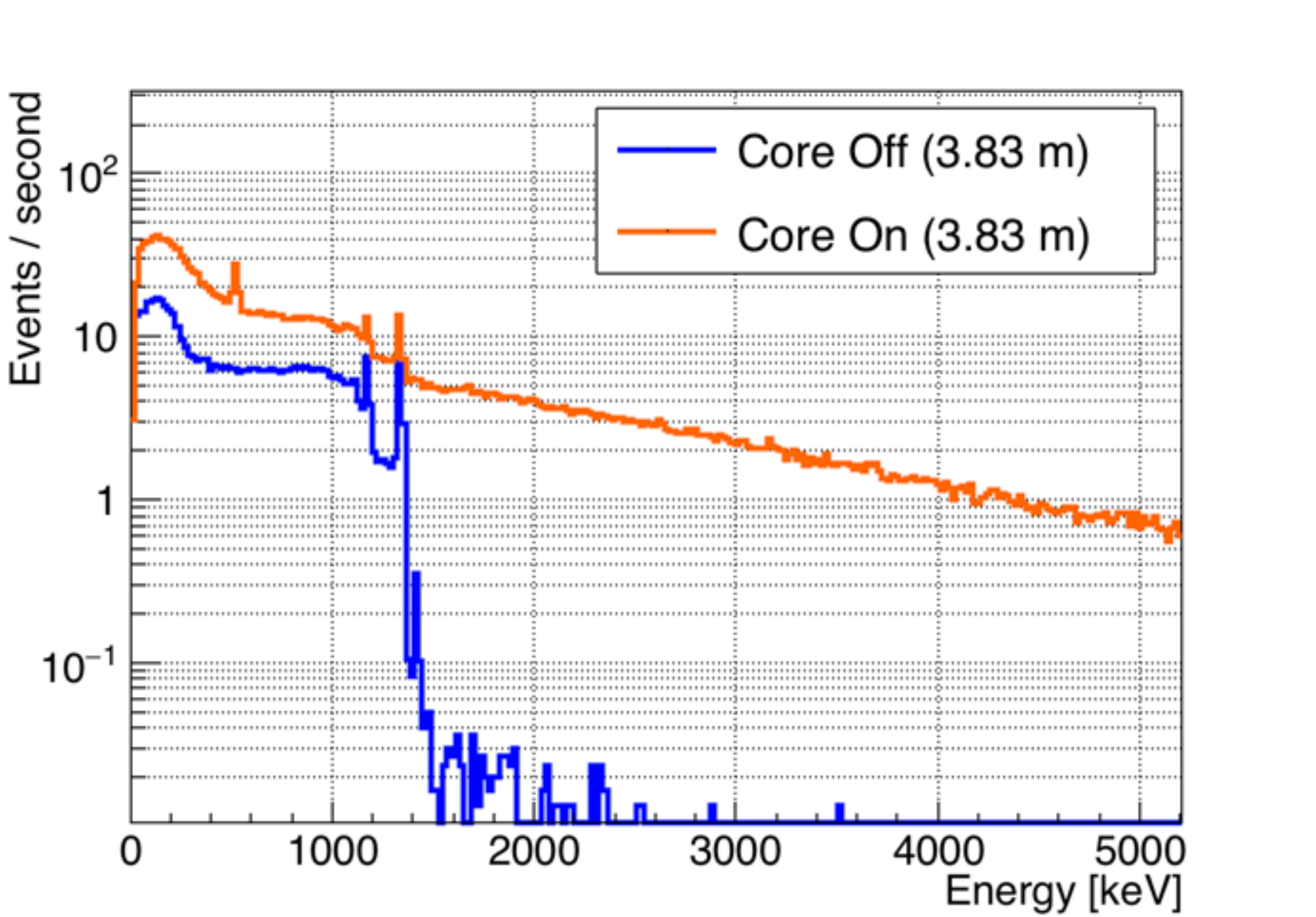}
\includegraphics[width=3.0in]{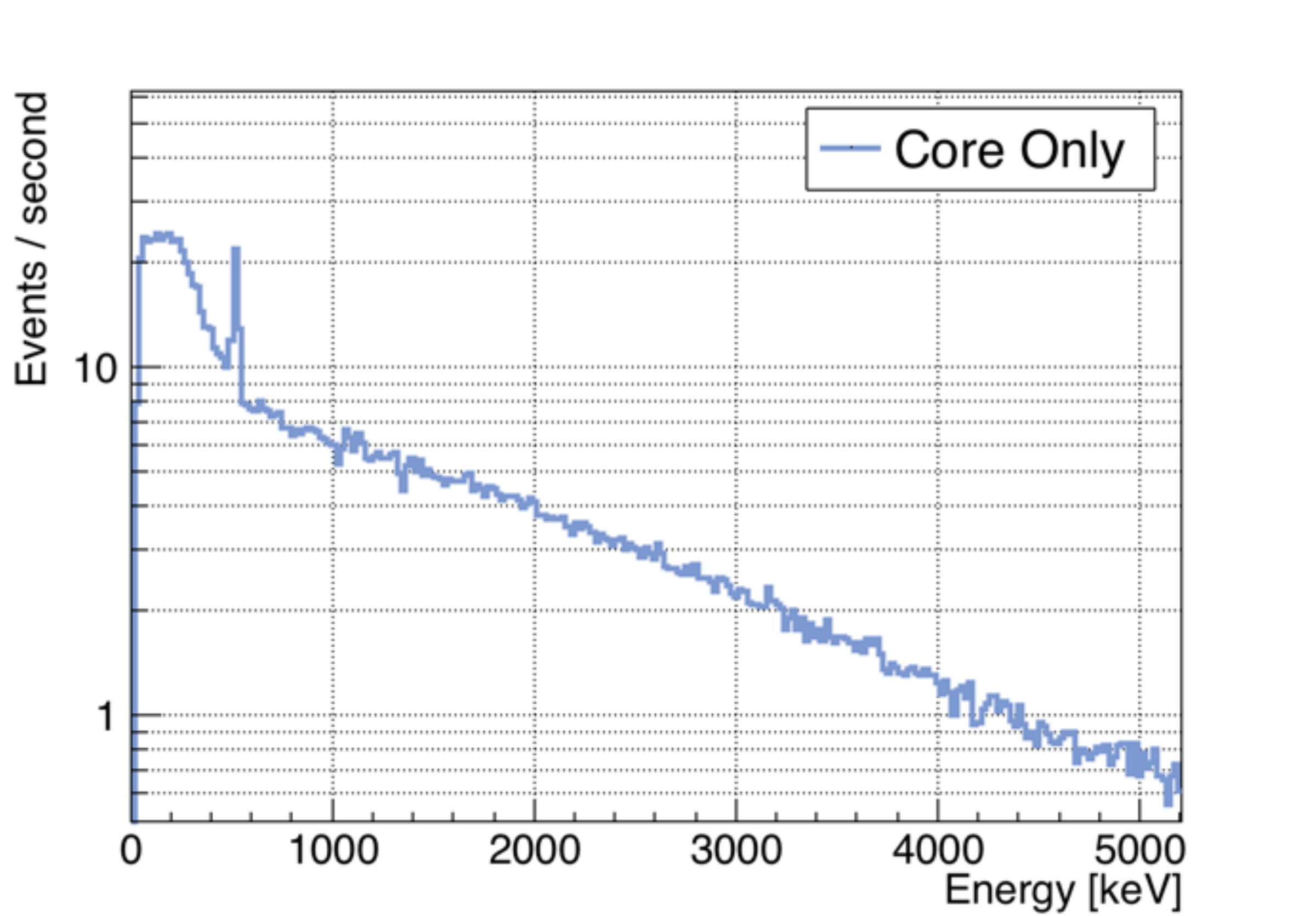}
\caption{
\footnotesize Full measured gamma spectrum in HPGe detector with the core (at distance 3.83~m) on and with the core off (left).  The core-off spectrum is subtracted from the core-on spectrum 
to get the ``core-only" spectrum (right) which can be compared directly to the simulation.  The main features of the core-off spectrum are the two Cobalt 60 gamma lines 
at 1173.2 and 1332.5 keV which come from activated stainless steel lining near the back of the experimental cavity.}
\label{fig:BkgSubtraction}
\end{figure}

Measured energy spectra are due to gammas 
produced in the reactor core and gammas from other sources (e.g., activated materials in the area). To properly compare simulations of the reactor core flux 
with measurements, the flux that is not coming directly from the reactor core must be subtracted.  To perform this subtraction, we use the energy spectra 
measured at a given core position, while the reactor is turned off.   An example of this subtraction is shown in Figure~\ref{fig:BkgSubtraction}.  It can be seen that after this subtraction, the only 
remaining spectral line is the 511\,keV electron-positron annihilation line.  Other lines present in the initial reactor core gamma energy spectrum are washed out due 
to gammas interacting in the water, graphite, and lead shielding between the core and detector.

\begin{figure}[ht]
\centering
\hspace{-5pt}
\includegraphics[width=3.0in]{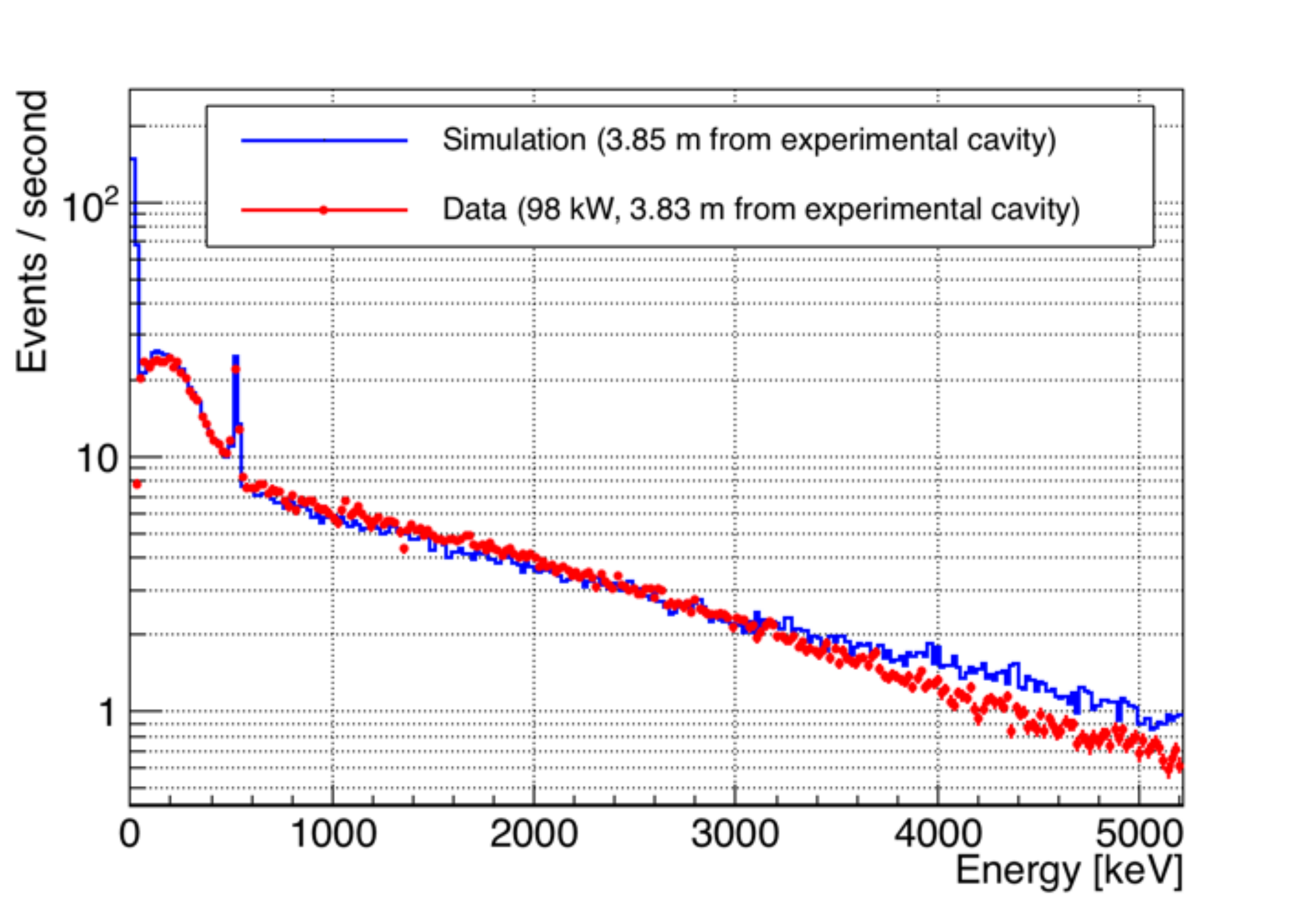}
\caption{
\footnotesize Comparison of reactor core gamma spectrum measured by the Canberra HPGe detector and the prediction from the GEANT4 simulation.  The GEANT4 
prediction is scaled to match the integrated event count of the data for the region of deposited energy greater than 100~keV in order to compare shape.  The large deviation in 
the region below 20\,keV is due to detector threshold effects not accounted for in the simulation.}
\label{fig:DataMCCompare}
\end{figure}

The different core positions were simulated in GEANT4, including a model of the lead shielding and the HPGe detector used 
to make the measurement (shown schematically in Figure~\ref{fig:Geant4HPGe}).   Each core position in 
simulation was systematically 2\,cm further from the detector than the corresponding position in the data, due to a correction made to the measured position 
after the simulation had been run.  For each core position, 
approximately $3\times10^9$ single gamma initial events were generated using 
the energy spectrum obtained from the MCNP core model described previously.  The deposited 
energy of gammas reaching the detector were compared 
to the background-subtracted data, as shown in Figure~\ref{fig:DataMCCompare}.  In this plot, the GEANT4 prediction was scaled to match the integrated event count of the data in the 
region of deposited energy greater than 100\,keV.  The energy resolution in the higher rate environment of the experimental cavity was dominated by pile-up 
effects resulting in a degradation of resolution with rate.  A resolution smearing was applied to the simulated results to account for this effect, determined using a comparison of width 
of the 511\,keV line between the simulation and the data.  A simple Gaussian fit was used to determine these line widths (ranging from about 4\,keV to 8\,keV), 
and the difference of the squares of these widths was used
to define a new Gaussian which was then applied as an event-by-event smearing to the simulated energy deposits.   The simulation and data matched quite well in shape, 
with about a 25\% deviation for the region above 3\,MeV.  The large deviation in 
the region below 20~keV is due to detector threshold effects not accounted for in the simulation.

\begin{figure}[ht]
\centering
\hspace{-5pt}
\includegraphics[width=3.0in]{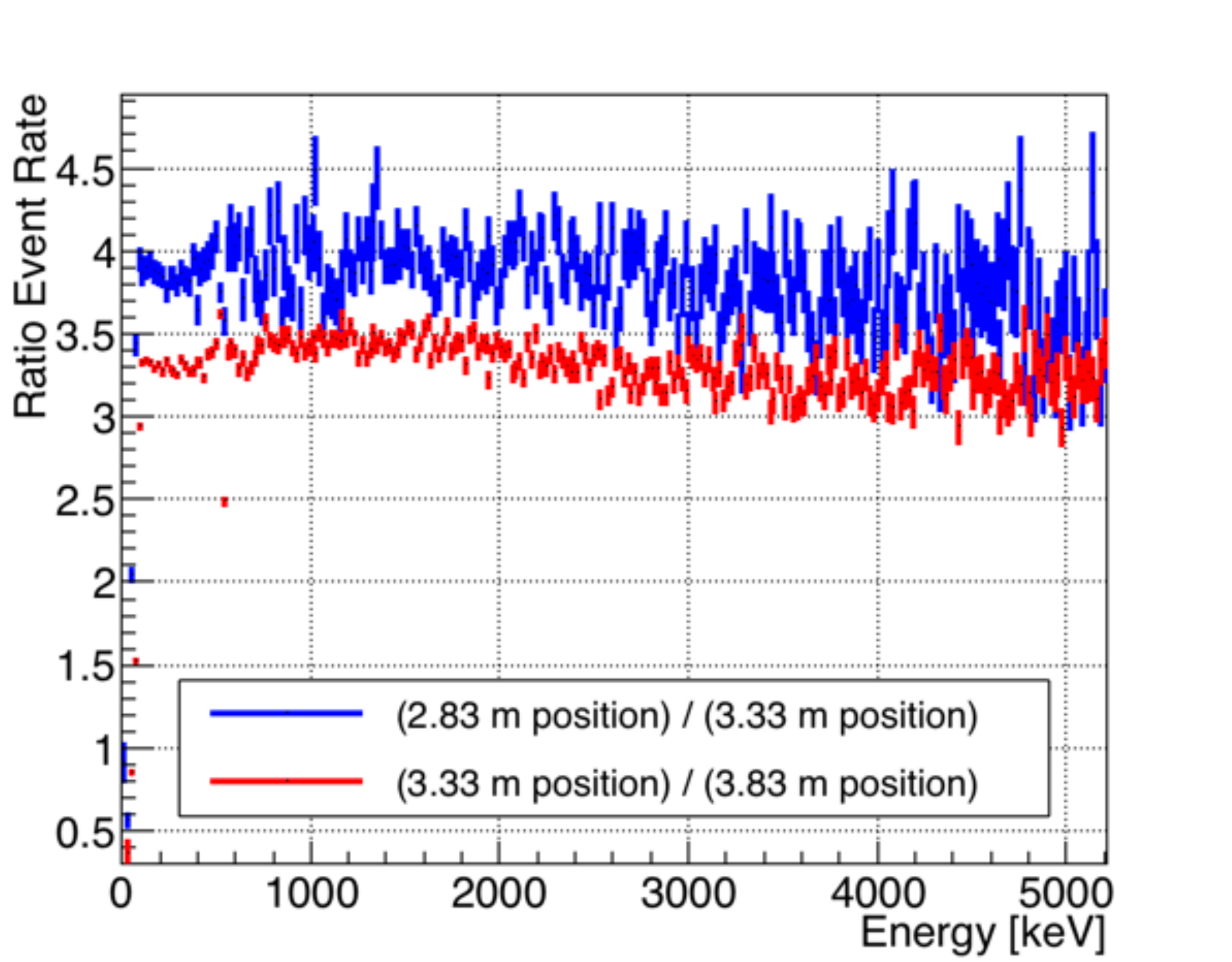}
\includegraphics[width=3.0in]{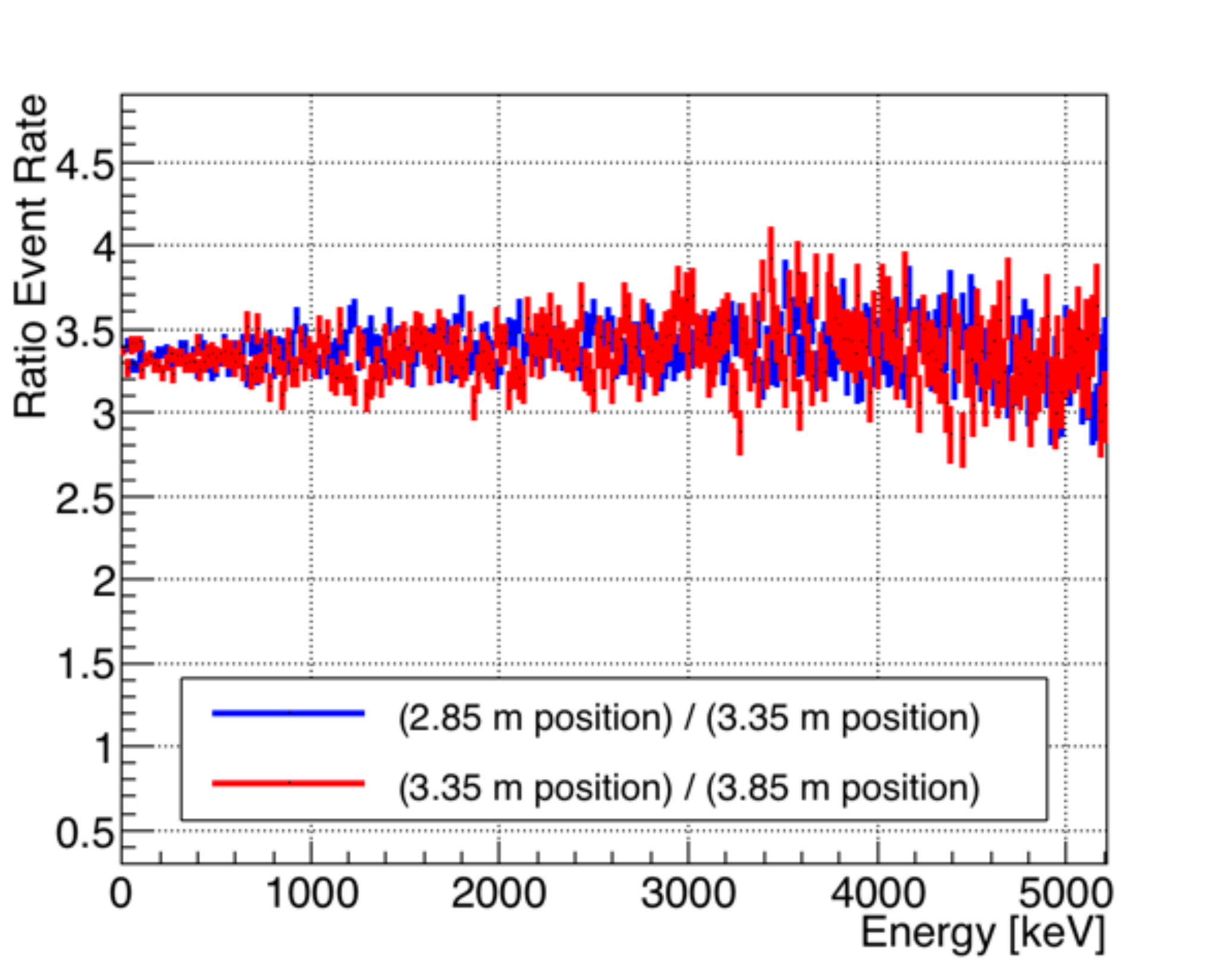}
\caption{
\footnotesize Event rate scaling as a function of deposited energy in the HPGe detector for different reactor core positions in the measured data (left) and simulation (right).  Error 
bars are statistical uncertainties only.}
\label{fig:RateScaling}
\end{figure}

We determined the scaling of event rate as a function of core position in both the measured and the simulated spectra by taking the ratio of background-subtracted 
spectra at each position.  This ratio is shown in Figure~\ref{fig:RateScaling}.  The rate is reduced by roughly a factor of 3.5 per 0.5\,m of increased distance from the core 
in both the data and the simulation.  
Because this scaling of rate with distance is reproduced to within about 10\% in the simulation, the scaling factor needed to translate the simulated result to a rate is taken 
to be constant regardless of the distance of the core to the detector.  We take as this scaling factor the ratio of the integrated event count of the data (in the region of deposited energy 
greater than 100\,keV) to the integrated weighted event count in the same region in the simulated result.  This scaling factor was applied to all further simulated results to obtain 
rate estimations. 

\section{Neutron Background Measurements}
\label{sec:Neutron}
\begin{figure}[ht]
	\begin{center}
		\includegraphics[height=2.1in]{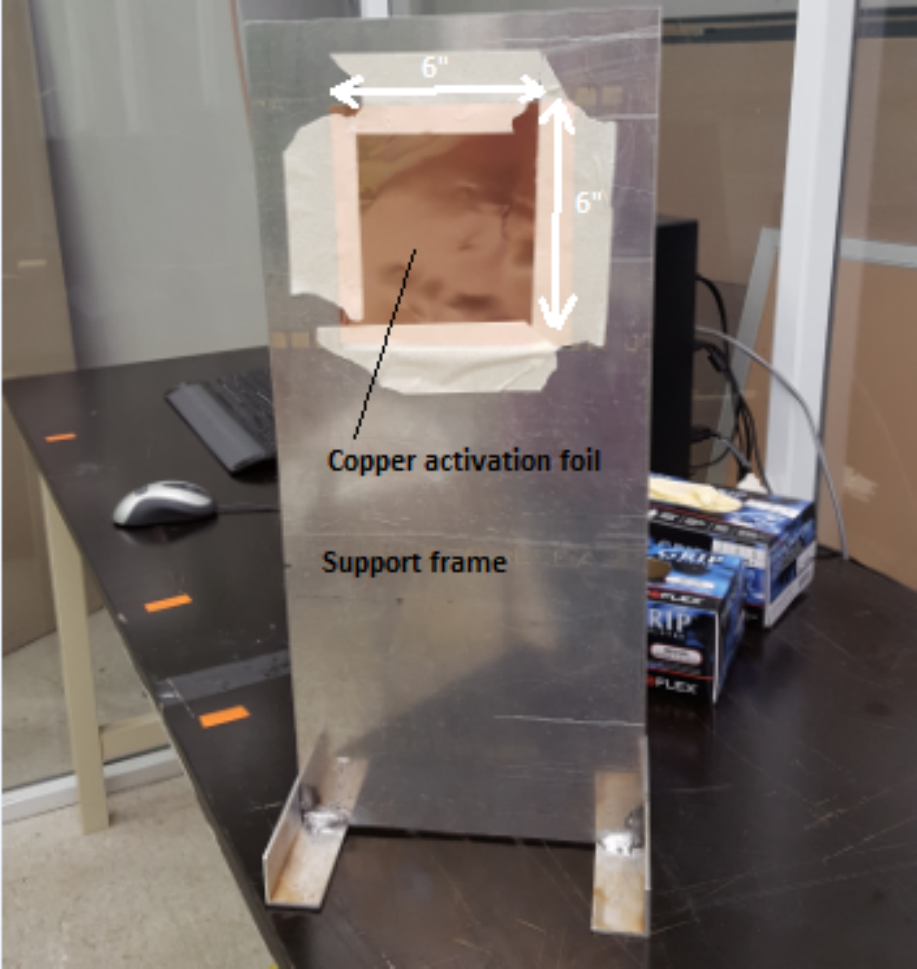}
		\includegraphics[height=2.1in]{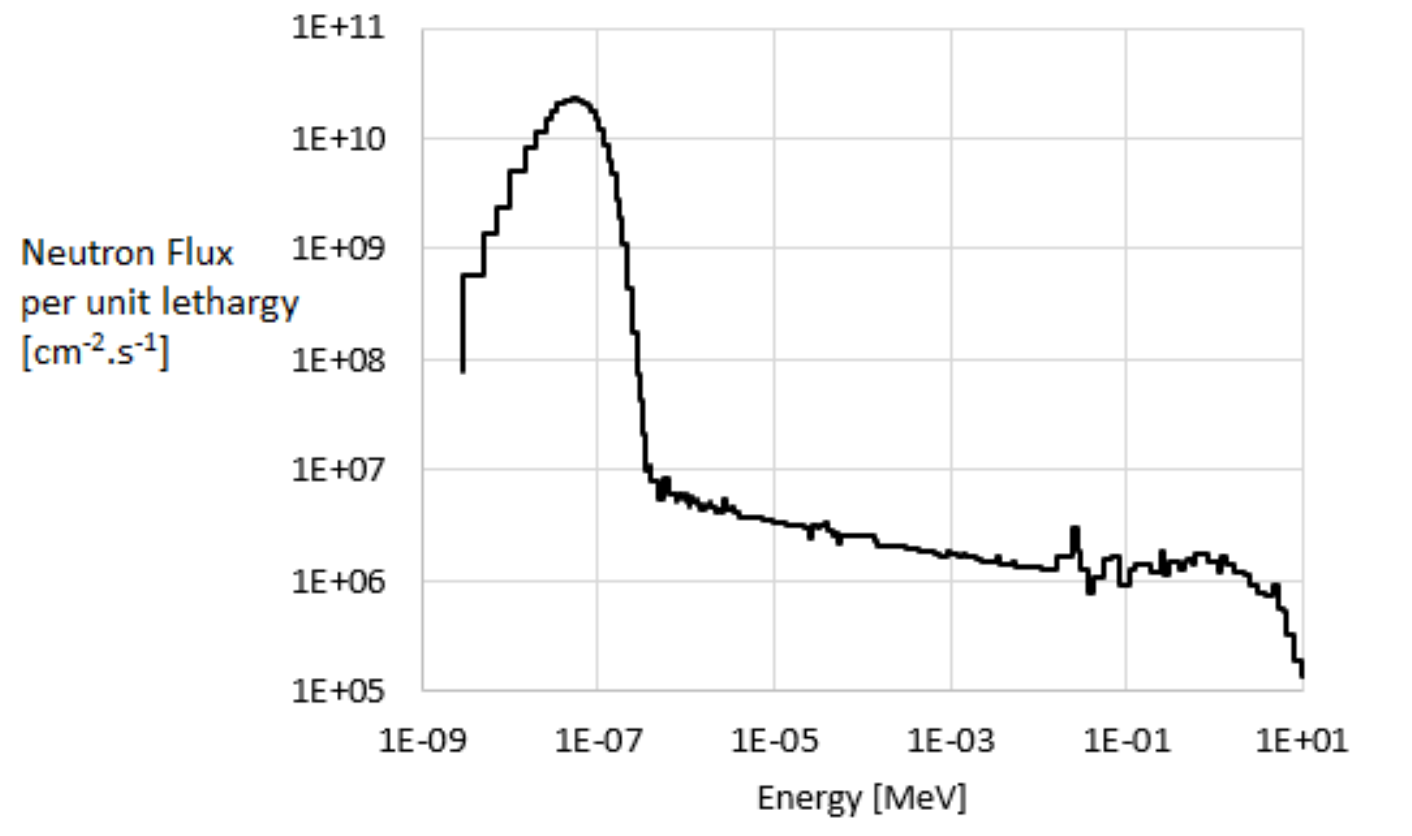}
		\caption{Photo of the $6\times6$\,in copper foil setup which was placed in the experimental cavity to be irradiated by neutrons(left).  Neutron spectrum inside the experimental 
		cavity at the surface of the graphite block facing the experimental cavity, as calculated by MCNP (right).  The neutron spectrum is bin-by-bin 
		normalized to $E_{Ave}/(E_2 - E_1)$ where $E_1$ and $E_2$ are the energy values at the respective bin edges and $E_{Ave}$ is the average of these bin edge values.}
		\label{figure:FoilHolder}
	\end{center}
\end{figure}

Due to administrative and safety constraints preventing deployment of traditional detectors, the background neutron measurement was thus far restricted to 
measurements needed for validation of the computational models.  The validation 
measurement was performed using a 6$\times$6~inch copper foil (see Figure~\ref{figure:FoilHolder}) that was activated by neutrons in the experimental cavity. The 
copper acts as an absorber of thermal neutrons 
and can be used to verify the integrated thermal neutron flux by measuring the activation of the foil after neutron exposure.  This measurement was performed with the 
core and experimental cavity in the configuration shown in Figure~\ref{fig:TCdiagram} (left) (i.e., without any additional shielding).

Using the MCNP simulation, the neutron spectrum at the surface of the graphite box facing the experimental cavity was determined using the 172 energy bin XMAS structure~\cite{HndbookNucVol1}. The 
calculated spectrum is shown in Figure~\ref{figure:FoilHolder} (right), with fluxes of $5.6(3) \times 10^{6}$\,cm$^{-2}$~s$^{-1}$ fast and $4.0(2) \times 10^{10}$\,cm$^{-2}$~s$^{-1}$ 
thermal neutrons, and was used in a subsequent activation analysis to obtain the capture cross-section of $^{63}$Cu.  This cross-section 
was then used in the Bateman equations~\cite{Bateman} to obtain the total thermal flux from the $^{64}$Cu measured activity. The activated $^{64}$Cu decays with a half life of 12.7\,hours 
via electron capture, beta and positron emission (0.5787\,MeV and 0.6531\,MeV respectively), and gamma emission (1.355\,MeV). The activity of the foil was 
measured by a HPGe detector at the NSC, resulting in a measured total thermal neutron flux of $5.8(3) \times 10^7$\,cm$^{-2}$s$^{-1}$.  The uncertainty on these fluxes include 
both statistical uncertainty from the model calculation and HPGe activation measurement as well as uncertainty on the core power calibration. 

\begin{figure}[ht]
	\begin{center}
		\includegraphics[height=2.1in]{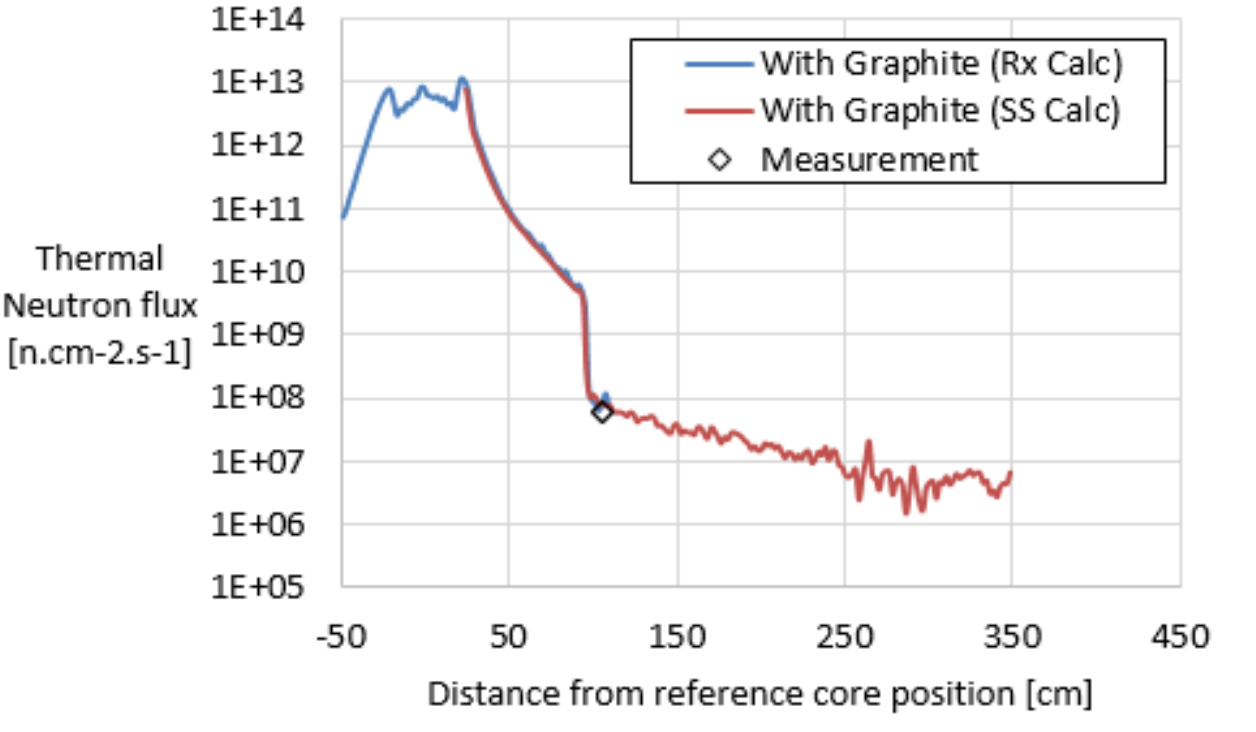}
		\caption{Thermal neutron flux as a function of distance from the reactor core center with only the graphite block between core and foil position.  The data point represents 
		the flux calculated using the measured activation of the copper foil at that  
		position.  The blue line (Rx Calc) indicates the flux calculated directly from the MCNP core model, while the red line (SS Calc) is this flux propagated 
		through the graphite block and into the experimental cavity using MCNP from the core model output.  }
		\label{figure:NeutronFluxProfile}
	\end{center}
\end{figure}

The measured thermal neutron flux together with a calculated thermal neutron flux profile is shown in Figure~\ref{figure:NeutronFluxProfile}.  The data point was consistent with the expected result from the  
simulation within 5\%, indicating that the neutron flux is predicted by the MCNP core model.

\section{Muon Background Measurements}
\label{sec:Muon}
 
Bolometric detectors with low thresholds are particularly vulnerable to large energy depositions from atmospheric muons. A typical solution for low rate experiments to this 
problem is to install such detectors deep underground, maximizing the overburden, and thus shielding of the detector.   For detecting higher rate processes, such as neutrino interactions near 
a nuclear reactor, a higher muon rate can be tolerated. The experimental cavity proposed for this experiment provides some overburden in the form of the high density concrete 
wall surrounding the reactor pool, as the cavity is located within this wall (see Figure~\ref{fig:TCdiagram}). This overburden has been characterized with regards 
to its muon shielding effectiveness by measurements described below.
 
Two polyvinyl-toluene scintillators (one smaller $1.75\times0.5\times0.375$\,in$^3$ panel on top and a larger $13.0\times3.0\times0.375$\,in$^3$ panel on bottom, separated 
by a 2\,in lead brick) were installed to trigger on muons. These counters were coupled to photomultiplier tubes via waveguides and were connected to front-end NIM electronics 
to produce a coincidence signal when both scintillators triggered above threshold within a 3\,ns window of each other.
Due to the high rate (and high energy) gamma environment, accidental triggers due to random coincidence were an experimental concern. To characterize 
the rate of such events, the signal of one scintillator was delayed arbitrarily to about 150\,ns, maintaining all other aspects of the experimental conditions. This setup 
was first exposed in the most radioactive location surveyed to get an upper limit on the rate of random coincidence events.  A 13\,hr run under these 
conditions showed no events passing the coincidence requirement, demonstrating that a subtraction correction for random coincidence would not be necessary.

\begin{figure}[ht]
\centering
\hspace{-5pt}
\includegraphics[width=3.0in]{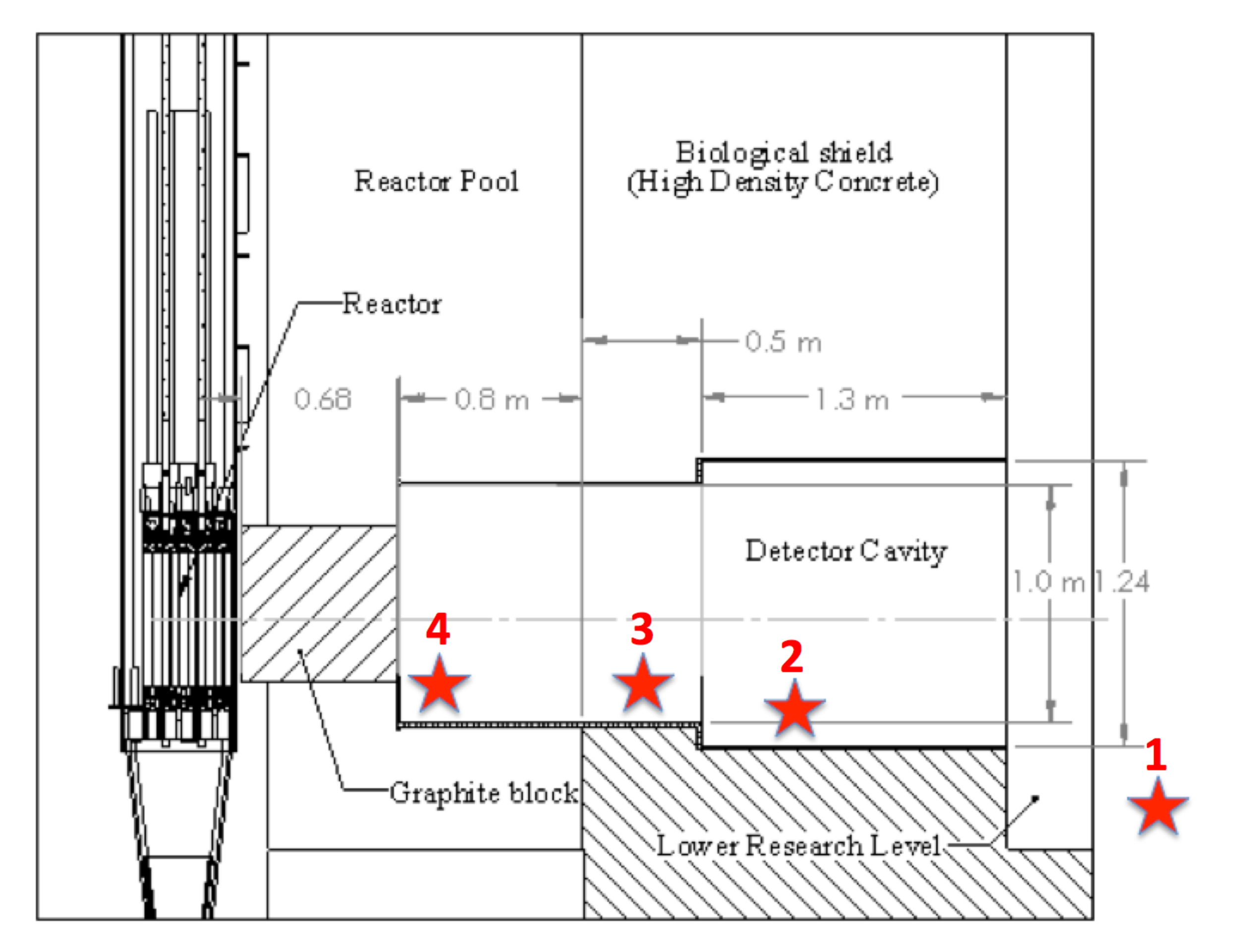}
\caption{
\footnotesize Approximate position of the muon scintillation detector in the experimental cavity for the four measurements made. }
\label{fig:MuonMeasureDia}
\end{figure}

Muon measurements were made with this setup at 5 locations. The first was made in a building adjacent to the reactor confinement building as a baseline measurement. This 
location has no effective overburden and was used as a open-sky muon baseline reference. With this setup, a rate of about 1\,$\mu$/min/cm$^2$ was measured at that 
location. The equipment was then moved to the lower research level of the reactor confinement building. The other measurements were made in the confinement building 
and all share the overburden of the 1\,m thick high density concrete roof. The next measurement, performed outside of the reactor pool wall, showed a 17\% reduction in 
muon rate compared to the baseline. The setup was then installed in the experimental cavity, inserted into 3 different positions in the cavity as shown in 
Figure~\ref{fig:MuonMeasureDia}.  The measured reduction in muon rate with respect to the open-sky baseline is given in Table~\ref{fig:MuonMeasure}. 

\begin{table}[ht]
\centering
\begin{tabular}{c c c}
\phantom{xx}Position \#\phantom{xx}  & \phantom{xx}Distance Into Cavity\phantom{xx} & \phantom{xx}Muon Rate Reduction\phantom{xx}  \\
\hline
1 & - & $17\pm6 \%$ \\
2 & 1.0~m & $50\pm3 \%$ \\
3 & 1.5~m & $57\pm6 \%$ \\
4 & 2.5~m & $47\pm4 \%$ \\
\end{tabular}
\caption{
\footnotesize The muon rate reduction with respect to the open-sky measurement.  Position number refers to the positions marked on the diagram in Figure~\ref{fig:MuonMeasureDia}.  The open-sky measurement was taken in a separate building at the NSC facility located at ground level. }
\label{fig:MuonMeasure}
\end{table}

These measurements show a 50\% reduction of muons incident upon the MINER detectors in the proposed experimental cavity. They also guide calculations 
of energy depositions that will be carefully considered in finalizing the geometry of the detectors. For a given volume/mass, one must make a trade-off between 
muon rate (determined effectively by a detector's horizontal cross-section) and energy deposition per muon (determined by detector's vertical dimension). 

\section{Rate Estimate With Shielding}
\label{sec:rateEstimate}

\begin{figure}[ht]
\centering
\hspace{-5pt}
\includegraphics[width=3.5in]{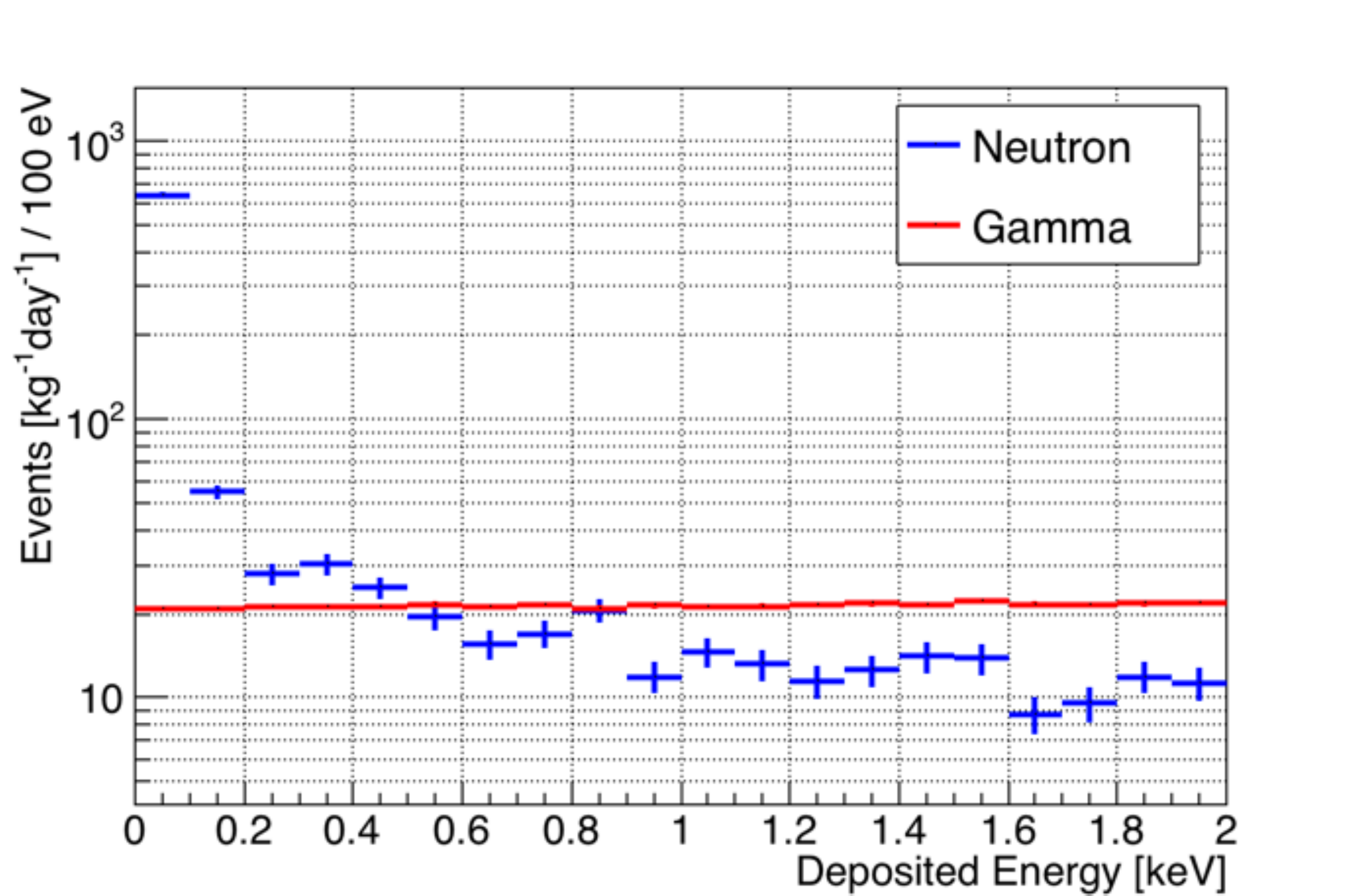}
\caption{
\footnotesize Energy deposited in the germanium detectors due to neutron and gamma backgrounds as estimated using the GEANT4 simulation and 
shielding design shown in Figure~\ref{fig:ShieldDesign1}.  The core source was located at the closest possible proximity to the experimental cavity.   
Uncertainties shown on this plot are statistical only.}
\label{fig:BkgEstimate}
\end{figure}

We then used the GEANT4 setup to estimate the backgrounds with 
a full preliminary shielding design, as shown in Figure~\ref{fig:ShieldDesign1}.  We generated approximately 
$3\times10^9$ gamma and $4\times10^9$ neutron events with the core at the closest 
possible proximity to the experimental cavity (at the face of the graphite block shown in Figure~\ref{fig:TCdiagram}).  The simulation included 4 germanium detectors and 4 silicon 
detectors, each represented as 100\,mm diameter, 33\,mm thickness cylinders, and backgrounds were assessed by determining the energy deposited in these volumes.  Rates were determined using 
the scaling obtained in the previous gamma and neutron measurements.  The 
resulting spectrum of energy deposited is shown in Figure~\ref{fig:BkgEstimate}.  

These rates are compatible with the target background rate of 100\,events/kg/day in the range of recoil energy between 
10 and 1000~eV$_{\rm nr}$ and optimization of the shielding configuration will further reduce the estimated rate. The event rate outside of this window was found to 
be approximately 30\,Hz summed in all detectors.  It should be noted that moving the core further 
away would drastically reduce the expected background, due to the addition of more water shielding between core and 
experiment, as well as the $r^{-2}$ reduction with distance.

\section{Summary and Future Prospects}
\label{sec:summary}

We have performed \textit{in situ} measurements and detailed simulations of expected backgrounds for the proposed MINER experiment with the goal of 
detecting {\cns}.  Simulations reproduce the measurements of thermal neutrons and gammas and were used to estimate the  
expected backgrounds with a full shielding designed to bring the backgrounds down to a level compatible with a measurement of the {\cns} signal in the MINER experiment.  This 
simulation has shown that it is indeed possible to reduce both neutron and gamma backgrounds to a level of about 100\,events/kg/day in the range of recoil energy between 
10 and 1000~eV$_{\rm nr}$ with the reactor core as close as about 2.3\,m.

Gamma measurements will be performed with increased lead 
shielding and a core at the closest position to the experiment.  Neutron measurements will be performed with other foil samples to provide multiple checks against the 
simulation, as well as with other detector technologies.  In parallel, optimization of the shielding configuration is being performed in both the GEANT4 and MCNP simulations 
to further reduce the neutron and gamma background.   
Simulation of muon-induced neutron backgrounds are now underway, but these are expected to be small in the experimental region of interest.

Based on measurements reported here and planned in the future, we are developing plans 
for \textit{in situ} monitoring of backgrounds for the MINER experiment. Several technologies are being 
considered including segmented, active liquid-scintillator shield, and a dedicated iZIP-type detector used 
by SuperCDMS~\cite{Brink:2006rs} that provides excellent nuclear recoil discrimination down to 1\,keV recoil energy.
Also, to further monitor and characterize neutron backgrounds, a $^6$Li doped scintillator detector for thermal 
neutrons, and a PTP-doped scintillator with neutron/gamma pulse shape discrimination for fast neutrons 
are being constructed at the Texas A\&M University Cyclotron Institute.

\section{Acknowledgements} 

The authors gratefully acknowledge the Mitchell Institute for Fundamental Physics and Astronomy for seed funding, as well as the Brazos HPC cluster 
at Texas A\&M University (brazos.tamu.edu) and the Texas Advanced Computing Center (TACC) at the University of Texas at 
Austin (www.tacc.utexas.edu) for providing resources that have contributed to the research results reported within this paper.   We also gratefully acknowledge the TAMU Nuclear Science Center 
for facilitating the MINER experiment and providing numerous resources and advice in the course of planning and developing this program.  R.M, R.H. and N.M acknowledge 
the support of DOE grant DE-SC0014036 in development of synergistic activities with the SuperCDMS collaboration.  G.V.R. acknowledges support by the U.S. Department of 
Energy, Office of Science, Office of Nuclear Science, under Award No. DE-FG02-93ER40773 and also support by the Welch Foundation (Grant No. A-1853).  J.W.W. acknowledges 
support from NSF grant PHY-1521105 and the Mitchell Institute for Fundamental Physics and Astronomy.  L.S. acknowledges support from NSF grant PHY-1522717.  B.D. acknowledges 
support from DOE grant DE-FG02-13ER42020.  

\section*{References}
\bibliographystyle{ieeetr}
\bibliography{MINERBackgrounds}

\end{document}